\documentclass[prd,
superscriptaddress,
onecolumn,
nofootinbib,
10pt]{revtex4-2}
\usepackage[colorlinks=true,
linkcolor=blue,
breaklinks=true,
urlcolor=magenta,
citecolor=blue]{hyperref}
\usepackage{graphicx,xcolor}% Include figure files
\usepackage{xtab,afterpage}
\usepackage{lipsum}
\usepackage{bm}% bold math
\usepackage{amssymb,amsmath,amsfonts, mathrsfs}

\begin{document}
\title{Revisiting $O(N)$ $\sigma$ model at unphysical pion masses and high temperatures. II. The vacuum structure and thermal $\sigma$ pole trajectory with cross-channel improvements}

% \title{Reinvestigating the vacuum structure and old problems of $O(N)$ $\sigma$ model with varying $m_\pi$ and temperature}



\author{Yuan-Lin Lyu}
\email[]{yllyu@stu.pku.edu.cn}
\affiliation{Institute for Particle and Nuclear Physics, College of Physics, Sichuan University, Chengdu  610065, People's Republic of China}
\affiliation{School of Physics, Peking  University, Beijing  100871, People's Republic of China}

\author{Qu-Zhi Li}
\email[]{liquzhi@scu.edu.cn}
\affiliation{Institute for Particle and Nuclear Physics, College of Physics, Sichuan University, Chengdu  610065, People's Republic of China}

\author{Zhiguang Xiao}
\email[]{xiaozg@scu.edu.cn}
\affiliation{Institute for Particle and Nuclear Physics, College of Physics, Sichuan University, Chengdu  610065, People's Republic of China}

\author{Han-Qing Zheng}
\email[]{zhenghq@pku.edu.cn}
\affiliation{Institute for Particle and Nuclear Physics, College of Physics, Sichuan University, Chengdu  610065, People's Republic of China}

%\thanks{}
% \date{\today}

\begin{abstract}
The effective potential of the $O(N)$ model at large $N$ limit is reinvestigated with varying pion mass and temperature. For large pion masses and high temperatures, we find the phenomenologically favored vacuum, located on the upper branch of the double-branched effective potential for physical $m_\pi$, moves to the lower branch and becomes no longer a local minimum but a saddle point. The existence and running of the tachyon pole are also  discussed. These phenomena indicate that the applicable energy range of $O(N)$ model is more and more limited as $m_\pi$ becoming larger and temperature going higher. With the effective coupling constant defined from the effective potential, the possible correspondence between the two branches of the effective potential and the two phases of the theory (distinguished by positive or negative coupling) is verified even with nonzero explicit  symmetry breaking and at finite temperature. Also, we generalize the $N/D$ modified $O(N)$ model to study the thermal trajectory of the $\sigma$ pole with the cross-channel contributions considered and find the thermal $\sigma$ pole trajectory resembles its counterpart with varying pion mass at zero temperature.     
\end{abstract}

\maketitle

\section{Introduction}

The well-known linear sigma model~\cite{Gell-Mann:1960mvl} is one of the most studied models in quantum field theory. With the help of large $N$ expansion, the $O(N)$ $\sigma$ model becomes exactly solvable at the leading $1/N$ order and demonstrates a variety of non-perturbative properties~\cite{Dolan:1973qd,Schnitzer:1974ji,Coleman:1974jh}. When considered as a toy model for low energy pion-pion scatterings, this model adopts a linear realization of the chiral symmetry $SU(2)_L \times SU(2)_R \simeq O(4)$, with the three pseudo-Goldstone bosons, i.e., $\pi$'s,  combined with another scalar-isoscalar particle $\sigma$ which acquires a vacuum expectation value (VEV), spontaneously breaking the $SU(2)_L\times SU(2)_R$ to $SU(2)_V$.  In contrast, the chiral perturbation theory
~\cite{Gasser:1983yg,Gasser:1984gg} lives solely in the broken phase of QCD and is formulated within the nonlinear realization of chiral symmetry~\cite{Coleman:1969sm,Callan:1969sn}. Though it can well
reproduce the low energy behavior of the $\pi\pi$ scattering, it
can not describe the phase transition from the chiral symmetry broken
phase to the symmetric phase explicitly with high temperatures. However,
taking  the advantage of the linear realization of the chiral
symmetry, it can be explicitly demonstrated in the $O(N)$ linear sigma model that 
chiral symmetry is restored at high temperature~\cite{Meyers-Ortmanns:1993dhx,Meyer-Ortmanns:1996ioo,Bochkarev:1995gi,Andersen:2004ae,Amelino-Camelia:1992qfe,Petropoulos:1998gt,Chiku:1998kd,Lenaghan:1999si,Fejos:2009dm}.

In our previous work in Ref.~\cite{Lyu:2024lzr}, by analyzing the leading $1/N$ order $\pi\pi$ thermal amplitude obtained in $O(N)$ model, the detailed $\sigma$ pole trajectory was drawn, and the asymptotic degeneracy of $\sigma$ and $\pi$'s were demonstrated at high temperature regardless of the different zero-temperature $m_\pi$ values.   It was also found that 
the pole structure and its behaviors with varying $m_\pi$ are consistent with the analyses on the $\pi\pi$ phase shifts extracted from the lattice QCD data (see, e.g., Refs.~\cite{Briceno:2016mjc,Gao:2022dln,Cao:2023ntr,Rodas:2023gma,Rodas:2024qhn,Rupp:2024tyh}).
Indeed, \emph{at the qualitative level}, the $O(N)$ $\sigma$ model can describe the behaviors of the lowest $f_0$ pole (dubbed $f_0(500)$ by PDG~\cite{ParticleDataGroup:2022pth}) in the $IJ=00$ channel~\cite{Lyu:2024lzr}.

For a long time, without any explicit symmetry breaking term in the Lagrangian, several interesting features and problems of $O(N)$ model have been pointed out and studied in different ways~\cite{Coleman:1974jh,Root:1974zr,Kobayashi:1975ev,Abbott:1975bn,Linde:1976qh,Bardeen:1983st,Bardeen:1986td,Nunes:1993bk}: (a) the two-branch structure of the effective potential and vacuum instability, (b) the existence of a tachyon for the first branch (within the dimensional regularization scheme), and (c) the Landau pole and triviality. In the large $N$ limit, the effective potential of the $O(N)$ model, $V(\phi)$, is double-valued hence having two branches~\cite{Kobayashi:1975ev,Abbott:1975bn}, which results directly from the non-perturbative effects incorporated by summing up infinite diagrams. For the first branch,  the scalar field can acquire a nonzero VEV, whereas for the second, the scalar field's VEV is always zero~\cite{Abbott:1975bn}. Either branch of $V(\phi)$ has an imaginary part in the large $|\phi|$ region~\cite{Abbott:1975bn}, and this phenomenon was claimed to be a signature for the intrinsic vacuum instability of $O(N)$ model~\cite{Linde:1976qh,Bardeen:1983st,Bardeen:1986td}. 
Actually, a possible way out is that, if adopting a UV cutoff regularization and viewing the $O(N)$ model as an effective theory, it has been shown that the effective potential can be real and convex~\cite{Nunes:1993bk}.   
Moreover, if choosing the vacuum to be on the first branch, it has been found that there is a tachyonic pole in the scattering amplitude at the leading order of the large $N$ expansion, which results from the fact that the local minimum on second branch has lower energy than the one on the first branch~\cite{Kobayashi:1975ev,Abbott:1975bn}. In general, the existence of tachyons could be a severe problem. In fact, the effective potential becomes everywhere complex at the next-to-leading $1/N$ order due to the tachyonic pole in the loop integral~\cite{Root:1974zr}. But if taking the $O(N)$ model as only an effective field theory below a certain cutoff scale, since the analytically continued $S$ matrix can still be considered reasonable in the region far away from the tachyon pole, the tachyon pole actually represents the ``cutoff'' scale of $O(N)$ model and thus can be dealt with carefully for higher $1/N$ order calculations~\cite{Ghinculov:1997tn,Fejos:2009dm}. Another problem is related to the divergence of the renormalized coupling constant at a finite energy scale, i.e., the Landau pole. The Landau pole found by perturbative calculations of the renormalization group equation (e.g., in QED),  can be ignored in some senses since the perturbation series are not suitable anymore when the coupling constant gets large.
Unfortunately, the $1/N$ expansion has no similar escape, and hence the existence of Landau pole might be thought as leading to the triviality of $O(N)$ model~\cite{Bardeen:1983st}. Though the triviality of four-dimensional $\phi^4$ scalar field theory is widely acknowledged (see for example, Refs.~\cite{Aizenman:1981du,Frohlich:1982tw,Luscher:1987ay,Luscher:1987ek,Luscher:1988uq,Luscher:1988gc,Hasenfratz:1987eh,Wolff:2009ke}), 
recently, a different point of view proposed in Ref.~\cite{Romatschke:2023sce} argues that the $O(N)$ model is nontrivial at the large $N$ limit. If the ``negative coupling phase'' of the theory is acceptable, then $O(N)$ model can be nontrivial in the IR region even when the UV cutoff tends to infinity.

This paper is devoted to revisiting the above problems when $O(N)$ symmetry is explicitly broken, particularly for unphysical pion masses and at finite temperatures. Since these problems already arise at the leading order of large $N$ expansion, it is reasonable to explore first the cases with nonzero $m_\pi$ and temperature at the same $1/N$ order.  We find that these problems become even more complicated than the situations with no explicit symmetry breaking and zero temperature. For large pion masses and high temperatures,  the phenomenologically favored vacuum, located on the upper branch of the double-branched effective potential for physical $m_\pi$, will move to the lower branch and become no longer a local minimum but a saddle point. 
Surprisingly, we also find that the possible correspondence~\cite{Linde:1976qh} between the double-branch structure of the effective potential and the two phases (positive and negative coupling) of $O(N)$ model still holds, even with nonzero pion mass and finite temperature.
These novel phenomena then indicate that the aforementioned problems are indeed related to the intrinsic non-perturbative nature of $O(N)$ model. We hope that the explorations of the vacuum structure with explicit symmetry breaking and finite temperatures, may shed more light on the understanding of the non-perturbative region of a quantum field theory like the $O(N)$ $\sigma$ model. Additionally, the $\sigma$ pole trajectory with varying temperature is obtained in the $N/D$ modified $O(N)$ model for the completeness of the calculation in Ref.~\cite{Lyu:2024lzr}, where the pole trajectory is determined only from the leading $1/N$ order thermal amplitude.

The outline is as follows.
In Sec.~\ref{sect:Veff_zeroT}, the $O(N)$ model effective potential at zero temperature is quickly reviewed. We carefully investigate the vacuum structure with varying $m_\pi$ and obtain a criterion for the existence of tachyon. 
In Sec.~\ref{sect:Veff_finiteT}, the finite-temperature effective potential is studied with different $m_\pi$ values and the running of tachyon pole is discussed. The latter indicates that the applicable energy range of $O(N)$ model is more and more limited as $m_\pi$ becoming larger and temperature going higher.
In Sec.~\ref{sect:eff_coupling}, the effective coupling constant is calculated from the effective potential at zero and finite temperature. The possible correspondence between the two branches of the effective potential and the two phases of the theory is verified even with nonzero explicit  symmetry breaking and at finite temperatures.
In Sec.~\ref{sect:sigma_traj}, the $N/D$ modified $O(N)$ model is generalized  to the situations with finite temperatures, and the $\sigma$  pole trajectory with varying temperature is obtained with the cross-channel effects included. 
Finally in Sec.~\ref{sect:conclusion}, we briefly summarize the main results and shortly discuss the further issues that could be explored in the future.

\section{The vacuum structure of $O(N)$ $\sigma$ model with varying $m_\pi$ at zero temperature\label{sect:Veff_zeroT}}

The $O(N)$ linear $\sigma$ model is described by the following Lagrangian,
\begin{align}\label{eq:Lagrangian ON model}
    \mathcal L = \frac{1}{2}\partial_\mu \phi_a \partial^\mu \phi_a -\frac{1}{2}\mu_0^2 \phi_a \phi_a - 
    \frac{\lambda_0}{8N}(\phi_a\phi_a)^2+ \alpha \phi_N\,,
\end{align}
where $\alpha\phi_N$ is an explicit symmetry breaking term and with $\alpha \neq 0$, pions becomes massive. In the broken phase, with $\langle \phi_N \rangle \equiv v \neq 0$ and the $O(N)$ symmetry broken into $O(N-1)$, there are $N-1$ pseudo-Goldstone bosons, $\pi_a \equiv \phi_a \,(a=1,\cdots,N-1)$, which can be used to  describe the pions for $N=4$, and we also define $\sigma\equiv \phi_N -v$. It is well known that $O(N)$ model is solvable at large $N$ limit, which  can be easily done with an auxiliary field trick~\cite{Coleman:1974jh}, i.e. by introducing another field $\chi$ to the Lagrangian (with an irrelevant constant omitted),
\begin{align}
    \mathcal L \to \mathcal L + \frac{N}{2\lambda_0}\left(\chi-\frac{\lambda_0}{2N}\phi_a\phi_a -\mu_0^2 \right)^2
    =  \frac{1}{2}\partial_\mu \phi_a \partial^\mu \phi_a +\alpha\phi_N +\frac{N}{2\lambda_0} \chi^2 -\frac{1}{2}\chi\phi_a\phi_a -\frac{N\mu_0^2}{\lambda_0}\chi\,.
\end{align} 
The effective potential of $O(N)$ model at the leading order $1/N$ expansion can  be standardly calculated as 
\begin{align}
    V(\phi,\chi) = -\alpha\phi_N -\frac{N}{2\lambda_0} \chi^2 +\frac{1}{2}\chi\phi^2 +\frac{N\mu_0^2}{\lambda_0}\chi   
    -\frac{i}{2} N \int \frac{\mathrm{d}^4 \ell}{(2\pi)^4} \log ( -\ell^2 +\chi - i 0^+ )\,,
\end{align} 
where $\phi_a$ and $\chi$ are now reduced to constant values,  $\phi^2\equiv \phi_a\phi_a$,  and $\langle \chi \rangle =m_\pi^2$ since the pion mass corrections are at higher $1/N$ orders. For brevity, we adopt the same notation for the classical fields and the original fields in the Lagrangian, which can be understood in the context. As in our previous paper Ref.~\cite{Lyu:2024lzr}, the renormalization conditions are chosen to be~\cite{Coleman:1974jh,Chivukula:1991bx}
\begin{align}
        \frac{\mu^2(M)}{\lambda(M)}&=\frac{\mu_0^2}{\lambda_0} +\frac{i}{2}  \int \frac{\mathrm{d}^4 \ell}{(2\pi)^4} \frac{1}{ \ell^2 + i 0^+} \,,
\label{eq:ren-cond1}\\
        \frac{1}{\lambda(M)} &=\frac{1}{\lambda_0}
-\frac{i}{2}  \int \frac{\mathrm{d}^4 \ell}{(2\pi)^4} \frac{1}{ (\ell^2  + i 0^+)(\ell^2 -M^2 + i 0^+)}\,.
\label{eq:ren-cond2}
\end{align}
\begin{figure}%[!htbp]
    \centering
    \includegraphics[width=0.4\textwidth]{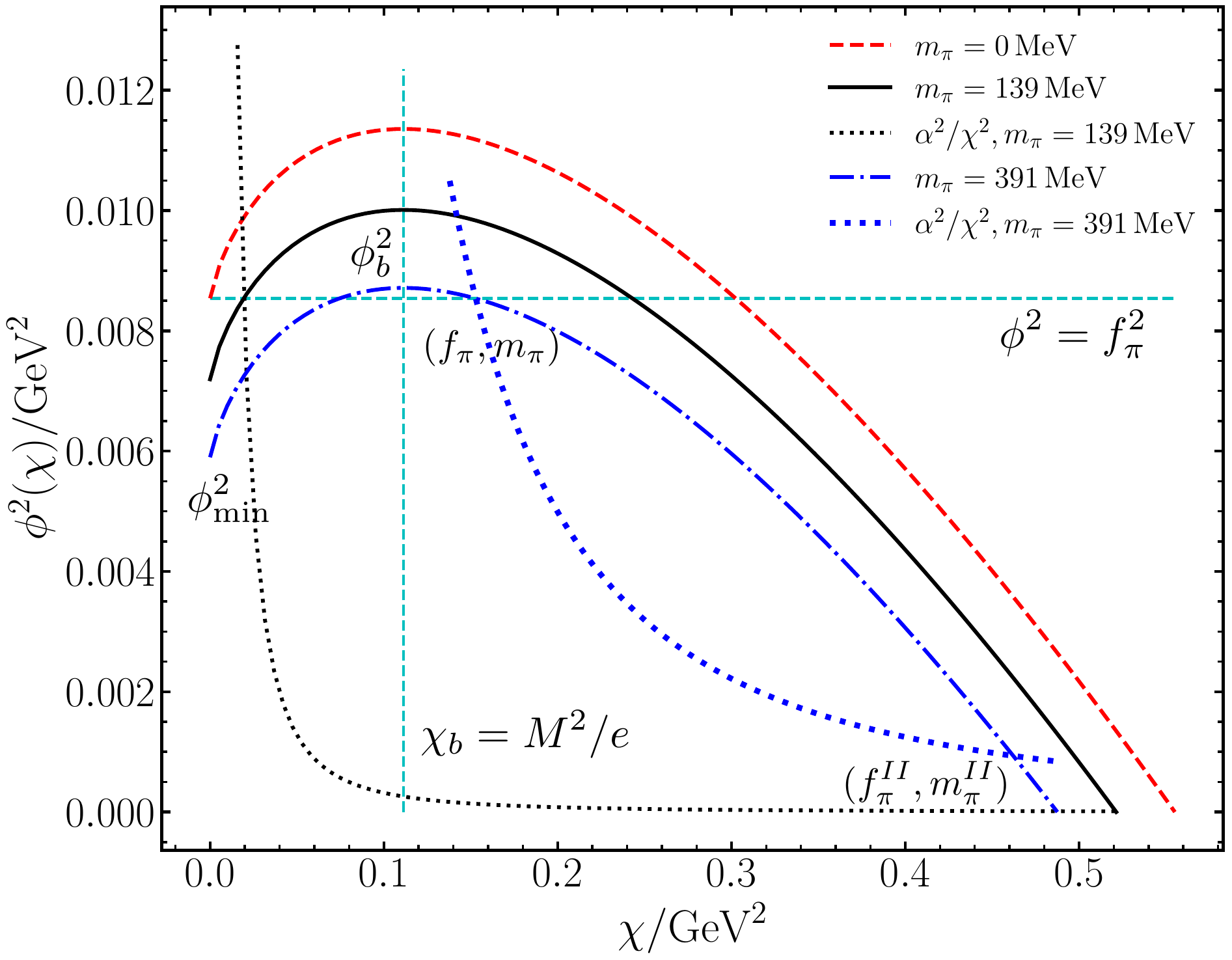}
    
    \caption{The typical behavior of $\phi^2(\chi)$ defined as Eq.~\eqref{eq:zeroT Gap eq 1}. The curves of $\phi^2(\chi)$ are drawn with $m_\pi=0$, $139$ and $391$ MeV respectively. The function $\alpha^2/\chi^2$ rearranged from Eq.~\eqref{eq:zeroT Gap eq 2} is also depicted for the last two cases with $\alpha = f_\pi m_\pi^2$ to demonstrate the other vacuum solution (at $v=f_\pi^{II}$ and $\chi=(m_\pi^{II})^2$) on the second branch of the effective potential. } 

    \label{fig:phi_sq of chi}   
\end{figure}
The effective potential can then be expressed in terms of the renormalized quantities as
\begin{align}
    V(\phi,\chi)&= -\alpha \phi_N +\frac{1}{2}\chi\phi^2 
    +\frac{N \mu^2(M)}{\lambda(M)}\chi -\frac{N}{64\pi^2}\chi^2\left( \log\frac{M^2}{\chi} +\frac{1}{2}\right)\,, 
    \label{eq:Veff zero T}
\end{align}
where we have chosen $M$ as the intrinsic scale of $O(N)$ model with $1/\lambda(M) =0$, i.e., the Landau pole. The vacuum is defined at the minimum of the effective potential, thus the conditions $\partial V/ \partial \chi =0$ and $\partial V/ \partial \phi_a =0$ result in the gap equations:
\begin{align}
    \phi^2 &=-\frac{2 N \mu^2(M)}{\lambda(M)}- \frac{N}{16\pi^2}\chi \log \frac{\chi}{M^2} \,,
\label{eq:zeroT Gap eq 1}\\
    \chi \phi_a &=0 \ (a<N)\,, \quad 
    \chi\phi_N =\alpha\,.
\label{eq:zeroT Gap eq 2}
\end{align}
Note that $V(\phi,\chi)$ is in general not identical to the effective potential of the original $O(N)$ model defined in Eq.~\eqref{eq:Lagrangian ON model}. In fact, the $O(N)$ model effective potential is recovered after eliminating the auxiliary field $\chi$ by taking $\partial V/ \partial \chi =0$ as a constraint condition~\cite{Coleman:1974jh,Abbott:1975bn}. Thus Eq.~\eqref{eq:zeroT Gap eq 1} will define $\chi(\phi^2)$ as a function of $\phi_a$ and the effective potential of $O(N)$ model, $V(\phi)\equiv V(\phi,\chi(\phi^2))$, is obtained.
However, Eq.~\eqref{eq:zeroT Gap eq 1} has two solutions of $\chi(\phi^2)$ which can be easily seen in Fig.~\ref{fig:phi_sq of chi}, where  the typical behavior of $\phi^2(\chi)$ from Eq.~\eqref{eq:zeroT Gap eq 1} is shown.
The two solutions of $\chi(\phi^2)$ are branched at $\chi = M^2/e \equiv \chi_b$ with $\chi^I(\phi^2) \le \chi_b \le \chi^{II}(\phi^2)$ and thus the effective potential also has two branches, $V^I(\phi)$ and $V^{II}(\phi)$, which is similar to the feature found in $O(N)$ model without explicit breaking~\cite{Kobayashi:1975ev,Abbott:1975bn,Linde:1976qh,Bardeen:1983st,Bardeen:1986td}. 
Moreover, the gap equations Eqs.~(\ref{eq:zeroT Gap eq 1}, \ref{eq:zeroT Gap eq 2}) also have  two solutions which can be seen
from the intersection points of the curves of $\phi^2(\chi^2)$ from
Eqs.~(\ref{eq:zeroT Gap eq 1}, \ref{eq:zeroT Gap eq 2}). 
For the solution corresponding to the left intersection point, $\chi$ will go
to zero as $\alpha$ tends to $0$, which is consistent with the mass of the usual
pions as pseudo-Goldstone bosons, i.e.  $m_\pi^2=\chi \to 0$  as $\alpha\to0$.
The solution for the right intersection point always sits on the second branch with
$v=f_\pi^{II}$ and $\chi=(m_\pi^{II})^2>\chi_b$, even for $\alpha\to0$ and thus can not describe the usual pion physics. From the
pseudo-Goldstone property of the left solution, we can express
$\mu^2(M)/\lambda(M)$ in terms of the usual $f_\pi$ and $m_\pi$ as follows. At zero temperature,  $\alpha = f_\pi m_\pi^2$, according to the partially conserved axial current (PCAC) relation, $\partial_\mu A_a^\mu=\alpha \pi_a$, where $A_a^\mu$ is the axial current, and the definition of the pion decay constant $f_\pi$, $\langle 0|A_a^\mu(x)|\pi_b\rangle=i p^\mu f_\pi e^{-ip\cdot x}\delta_{ab}$. Then with Eq.~\eqref{eq:zeroT Gap eq 2} and $\chi=m_\pi^2$, we have $v \equiv \phi_N= f_\pi$. And from Eq.~\eqref{eq:zeroT Gap eq 1}, the value of $\mu^2(M)/\lambda(M)$ is determined, $\mu^2(M)/\lambda(M) = - f_\pi^2/(2N) - m_\pi^2 \log (m_\pi^2/M^2)/(32\pi^2) $. Thus Eq.~\eqref{eq:zeroT Gap eq 1} becomes
\begin{align}
    \phi^2 =f_\pi^2 + \frac{N}{16\pi^2}\left(
    m_\pi^2 \log \frac{m_\pi^2}{M^2}-\chi \log \frac{\chi}{M^2}\right)\,.
\end{align}

From Eq.~\eqref{eq:zeroT Gap eq 1}, $\phi^2(\chi)$ has a maximum $\phi^2_b$ at $\chi = \chi_b$. It then follows that $V(\phi)$ has a nonzero imaginary part in the region $\phi^2>\phi^2_b$, since in that region there will be no real $\chi$ solutions and $\chi$ has to be complex~\cite{Coleman:1974jh}.
Additionally, Eq.~\eqref{eq:zeroT Gap eq 1} shows that for $V^I(\phi)$ there is another region $\phi^2 < \phi^2_\text{min} \equiv \phi^2(\chi=0)$ where it becomes complex. This unphysical region appearing in $V^I(\phi)$ is very similar to the case that one  encounters in the one-loop perturbative calculation of the effective potential in the broken phase (see textbooks, e.g. Refs.~\cite{Peskin:1995ev,Weinberg:1996kr}). Fortunately, it should be considered harmless to the theory because the effective potential can still be defined as real and convex when $\phi^2<\phi^2_\text{min}$ by choosing the state that satisfies $\langle \phi_a\rangle=\phi_a$ to be a proper linear combination of the states corresponding to other field expectation values where $V^{I}(\phi)$ is real~\cite{Weinberg:1996kr}. 
However, the above argument does not apply to the former region, i.e. $\phi^2>\phi^2_b$; thus, the nonzero imaginary part of $V(\phi)$ along that region and the double-branch feature are actually rooted in the non-perturbative effects captured by the large $N$ expansion. 
\begin{figure}%[!htbp]       
    \centering
    \includegraphics[width=0.4\textwidth]{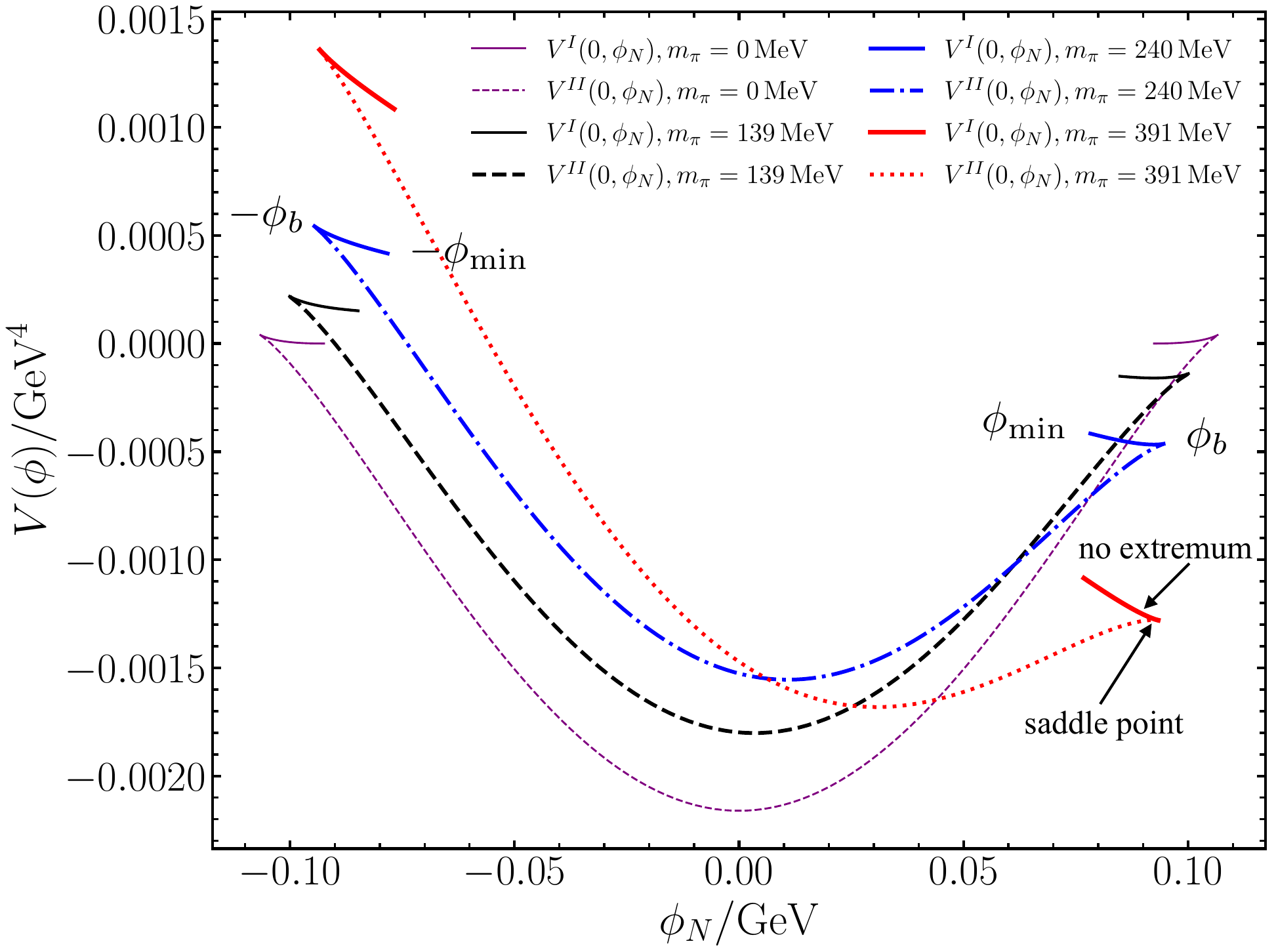}
    \quad
    \includegraphics[width=0.4\textwidth]{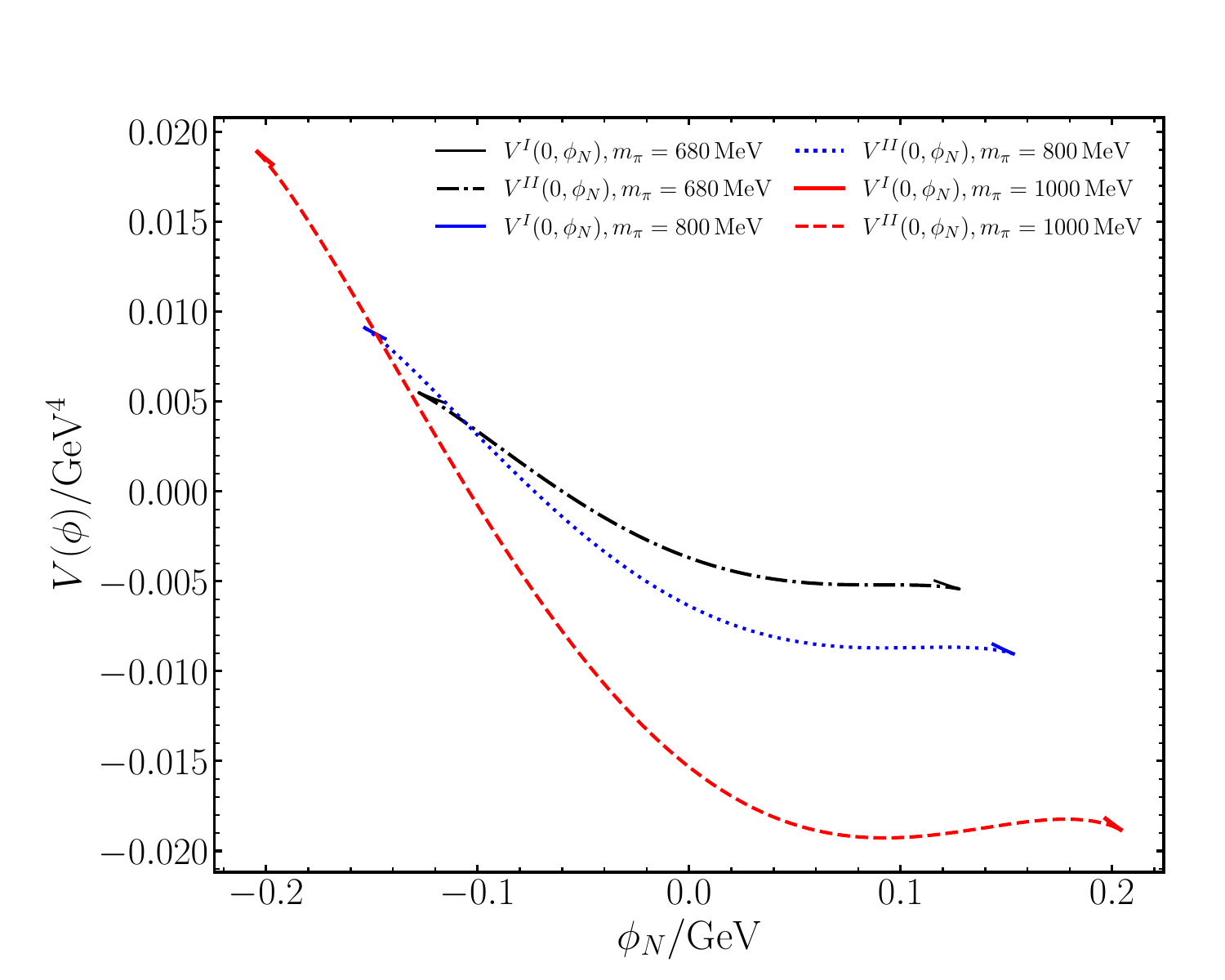}
    \caption{Effective potential with varying $m_\pi$ at zero temperature. For clearness of the presentation, the plots are limited to a ``slice'' of the effective potential along the $\phi_N$ direction with $\phi_a(a<N)$ set to their vacuum expectation values $\phi_a =0(a<N)$. The plots only show the region where the effective potential is real.  The cases correspond to different  pion mass values $m_\pi= 0$, $139$, $240$, $391$, $680$, $800$ and $1000$ MeV respectively.}   
    \label{fig:Veff varying mpi zeroT}  
\end{figure}

The double-valued effective potentials  with different $m_\pi$ values are plotted in Fig.~\ref{fig:Veff varying mpi zeroT}. For numerical calculations, we always set $N=4$ and $f_\pi=92.4$ MeV. The intrinsic scale of $O(N)$ model in leading $1/N$ order calculations is chosen as $M = 550$ MeV.

\subsection{The local stability of the effective potential}

In quantum field theory, the true vacuum state is  defined to be at the absolute minimum of the effective potential. The other local minima of $V(\phi)$  correspond to metastable vacuum states (false vacua). Moreover, the local maxima or saddle points are unstable configurations which cannot be viewed as stationary states.  Generally, for a  convex effective potential, all the vacuum states are among the solutions of the gap equations, i.e. Eqs.~(\ref{eq:zeroT Gap eq 1}, \ref{eq:zeroT Gap eq 2}) for $O(N)$ model. 
In Fig.~\ref{fig:Veff varying mpi zeroT}, as $m_\pi$ increases, the local minimum solution with $v=f_\pi$ and $\chi = m_\pi^2$  gradually moves from the first (upper) branch $V^I(\phi)$ to the second (lower) branch $V^{II}(\phi)$ and turns into a saddle point, since $\left.\partial^2 V(\phi)/ \partial\phi_N^2 \right|_v <0$ and $\left.\partial^2 V(\phi)/ \partial\pi_a^2 \right|_{\pi_a =0} = \chi >0$.\footnote{It is worth mentioning that the $\sigma$ particle mass, $m_\sigma$, is obtained from the pole position of the $IJ=00$ channel $\pi\pi$ scattering amplitude at the leading $1/N$ order~\cite{Lyu:2024lzr}, which is identical to the pole position in the leading $1/N$ order $\sigma$ propagator, and is generally not the same as the value of $(\partial^2 V/\partial\sigma^2|_{\sigma=0})^{\frac{1}{2}}$ due to the infinite sum of Feynman diagrams in the large $N$ calculation of the $\sigma$ propagator.} Then for larger pion masses there is no extremum solution on the first branch of the effective potential. 

\begin{figure}%[!htbp]       
    \centering
    \includegraphics[width=0.4\textwidth]{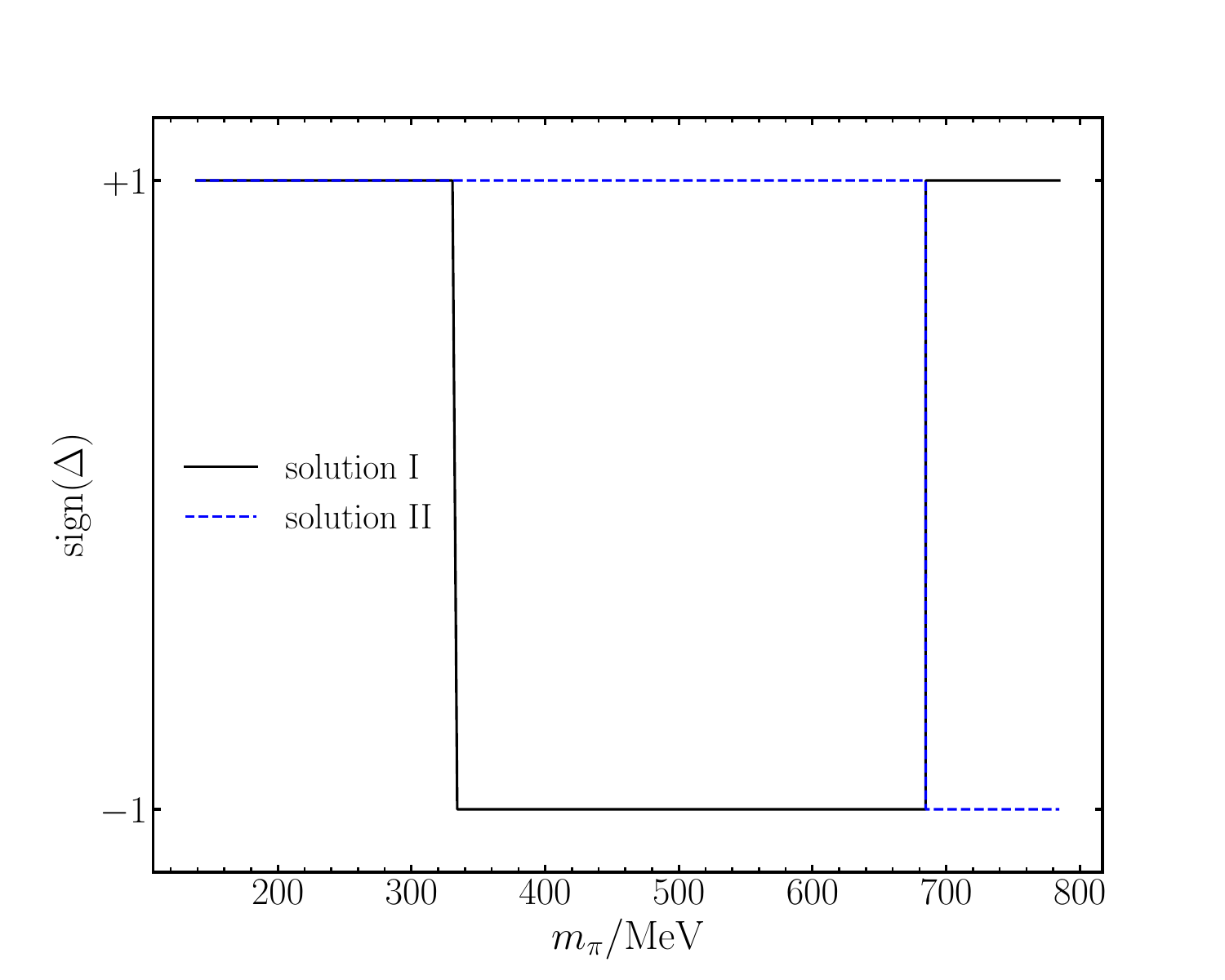}
    \caption{The sign of the criterion function $\Delta$ for the solution of gap equations being a local minimum of effective potential. Solution I denotes the vacuum at $v=f_\pi$ with $\chi=m_\pi^2$, which is on $V^I(\phi)$ at first and moves to $V^{II}(\phi)$ when $m_\pi>M/\sqrt e$. Solution II is the other vacuum always lying on $V^{II}(\phi)$.} 
    \label{fig:Veff criterion minimum zeroT}    
\end{figure}

\begin{figure}
    \centering
    \includegraphics[width=0.4\textwidth]{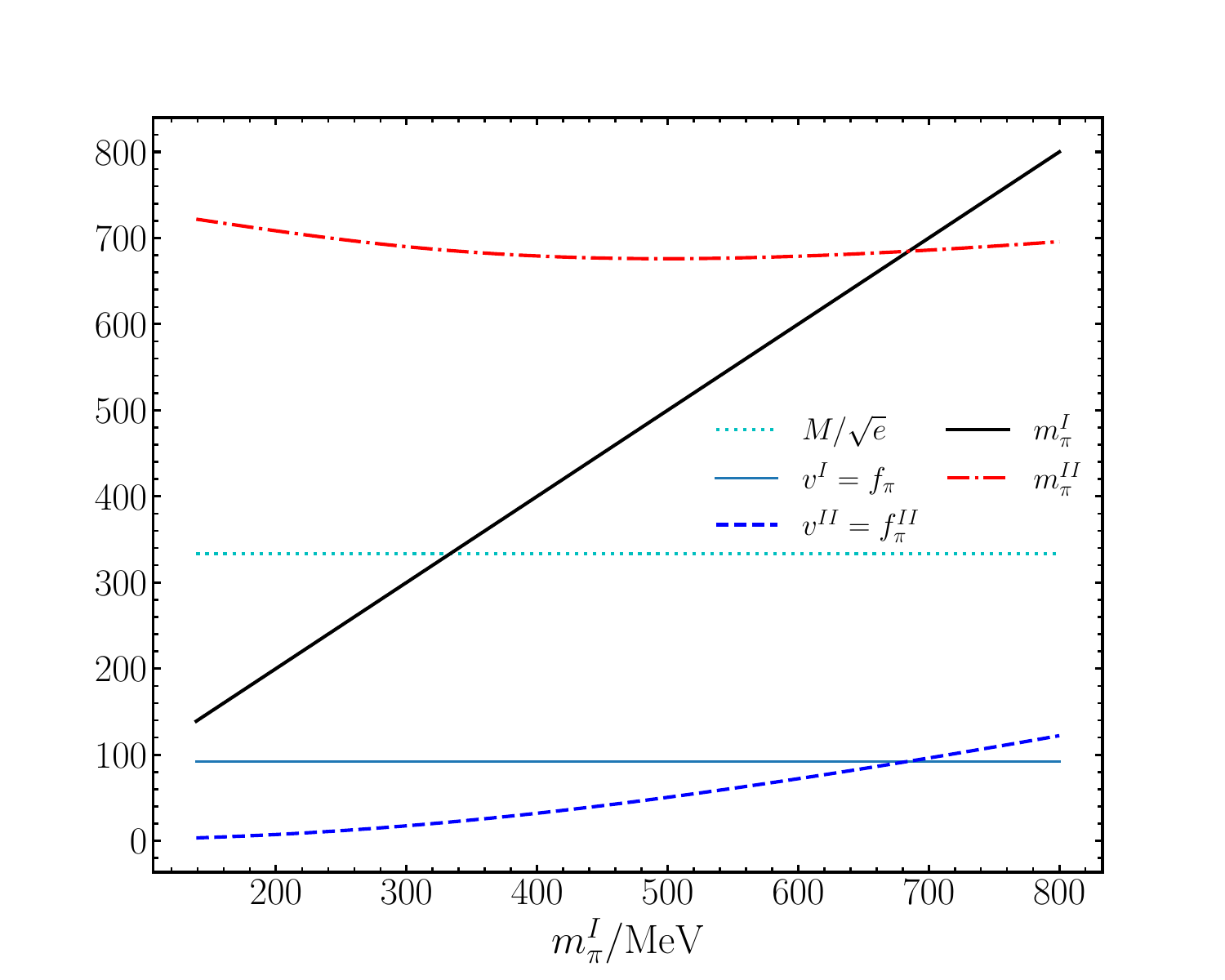}

    \caption{$v^{I/II}$ and $m_\pi^{I/II}$ with varying $m_\pi^I$ at zero temperature. The superscript I/II corresponds to solution I/II as defined in the main text.}
    \label{fig:vT_mpiT_varying_mpi_zeroT}  
\end{figure}

In the following we will show the details about how the vacuum solution becomes a saddle point when it moves to the second branch. 
Firstly, according to Eqs.~(\ref{eq:zeroT Gap eq 1}, \ref{eq:zeroT Gap eq 2}) and Fig.~\ref{fig:phi_sq of chi}, it is straightforward to demonstrate that when seeking for a vacuum at $v=f_\pi$ with $m_\pi<\sqrt{\chi_b}$, the solution lies on $V^I(\phi)$; while with $m_\pi>\sqrt{\chi_b}$, the solution is on $V^{II}(\phi)$. Besides this solution, there is always another vacuum located on $V^{II}(\phi)$.  Expanding the effective potential Eq.~\eqref{eq:Veff zero T} around the solutions of the gap equations, we have
\begin{align}\label{eq:Veff at neighbourhood of vac}
     V(\phi)   = V(0,v)+ \frac{1}{2}\chi \bar\phi^2 +\frac{1}{2}\left( \left.\frac{\partial \chi}{\partial \phi_N}\right|_v v +\chi \right)\sigma^2
    +\mathcal{O}\left((|\phi|-v)^3 \right) \,,
\end{align}
where  $\bar\phi^2 = \sum^{N-1}_{a=1} \phi_a^2$ and $ \left.\partial \chi / \partial \phi_N  \right|_v =  32\pi^2 v/[N(\log(M^2/\chi) -1)]$, which can be obtained from Eq.~\eqref{eq:zeroT Gap eq 1}.  In the above expansion, we have used the relation $\partial V(\phi)/\partial \bar\phi^2 = \partial V(\phi,\chi)/\partial \bar\phi^2 + (\partial V(\phi,\chi)/\partial\chi ) (\partial \chi/\partial \bar\phi^2)$ and Eq.~\eqref{eq:zeroT Gap eq 1}, i.e. $\partial V (\phi,\chi)/\partial \chi = 0$.
Considering that $\chi=m_\pi^2>0$ is always satisfied, the criterion for the solution of vacuum to be a local minimum is
\begin{align}\label{eq:criterion of Veff extremum zeroT}
    \Delta\equiv\frac{32\pi^2v^2/N}{\log(M^2/\chi) -1} +\chi >0\,.
\end{align}
The sign of the criterion function $\Delta$ for the two solutions (solution I/II, solution I denotes the vacuum at $v=f_\pi$ with $\chi=m_\pi^2$ and solution II is the other vacuum solution to the gap equations at $v=f_\pi^{II}$ with $\chi = (m_\pi^{II})^2$) are shown in Fig.~\ref{fig:Veff criterion minimum zeroT}. 
When $m_\pi<\sqrt{\chi_b}$, both solutions are indeed local minimums as expected. While $\sqrt{\chi_b} < m_\pi < (32\pi^2v^2/(Ny))^{1/2}$ (with $y e^y = \frac{32\pi^2 e v^2}{N M^2} $), the criterion Eq.~\eqref{eq:criterion of Veff extremum zeroT} is unsatisfied, resulting in  a saddle point of $V(\phi)$. 
For larger pion masses, $m_\pi \ge (32\pi^2v^2/(Ny))^{1/2}$, the two solutions I and II will hit  and move across each other (which can be seen in Fig.~\ref{fig:Veff varying mpi zeroT} and more clearly in Fig.~\ref{fig:vT_mpiT_varying_mpi_zeroT}), where solution I turns into a local minimum and solution II becomes a saddle point.\footnote{Actually, this feature can be analytically verified, using the fact that if there are two different solutions of the gap equations, the derivative of the function $\phi^2(\chi)-\alpha^2/\chi^2$  should have opposite signs for the two solutions, i.e. 
$\frac{\mathrm{d}}{\mathrm{d}\chi} ( \phi^2(\chi) -\alpha^2/\chi^2)|_{\chi=m_\pi^2}
\frac{\mathrm{d}}{\mathrm{d}\chi} ( \phi^2(\chi)-\alpha^2/\chi^2)|_{\chi=(m^{II}_\pi)^2}<0$. Noticing that for solution I and II, $\frac{\mathrm{d}}{\mathrm{d}\chi} \left( \phi^2(\chi)-\alpha^2/\chi^2\right) = \frac{N}{16\pi^2\chi} \left(\log(M^2/\chi)-1 \right)\left(\frac{32\pi^2v^2/N}{\log(M^2/\chi)-1}+\chi\right)$, where the last factor is exactly the criterion function $\Delta$ in Eq.~\eqref{eq:criterion of Veff extremum zeroT}, thus the local stability criterion can not be satisfied at the same time for solution I and II when they both lie on the second branch of the effective potential.  
}
By definition, the saddle point of the effective potential can not be chosen as the vacuum (or even a false vacuum). In fact, this situation does occur for solution I when $\sqrt{\chi_b}< m_\pi < (32\pi^2f_\pi^2/(Ny_0))^{1/2}$ (with $y_0 e^{y_0} = \frac{32\pi^2 e f_\pi^2}{N M^2} $) and for solution II when $m_\pi > (32\pi^2f_\pi^2/(Ny_0))^{1/2}$, if it is on the second branch, as shown in Fig.~\ref{fig:Veff criterion minimum zeroT}.

Notably, when using the $O(N)$ linear $\sigma$ model in the calculation of low-energy pion scattering~\cite{Lyu:2024lzr}, the phenomenologically favored vacuum is exactly the one on the first branch, $V^I(\phi)$, though there is another ``hidden" vacuum state with lower energy on the second branch.\footnote{This may be acceptable since the decay rate of the false vacuum is approximately $\propto e^{-\mathcal O (N)}$ (see, e.g. Refs.~\cite{Coleman:1977py,Bardeen:1983st}).} The phenomena of the local stability of the $O(N)$ model effective potential discussed above may imply that for large unphysical $m_\pi$, the acceptable vacuum could only be chosen at the second branch and also there should be a sharp transition for the choice of the vacuum state, after the local minimum moves from $V^{I}(\phi)$ to  $V^{II}(\phi)$ and becomes a saddle point. 

\subsection{The existence of a tachyon}

It has been complained for a long time that there is a tachyon found in the pion-pion scattering amplitude in $O(N)$ model at leading order $1/N$ expansion~\cite{Coleman:1974jh,Abbott:1975bn,Chivukula:1991bx,Ghinculov:1997tn}. Generally, one would argue that the existence of tachyon means the instability of the vacuum. However, things get more complicated in $O(N)$ model at large $N$ limit, since the effective potential is double-valued and may have two very distinct solutions with respect to the choice of the vacuum. Therefore it should be carefully examined that whether there is still a tachyon pole in each of the gap equation solutions when there is an explicit symmetry breaking term.

In $O(N)$ model at leading $1/N$ order, the $\pi\pi$ scattering amplitude~\cite{Coleman:1974jh} is 
\begin{align}
    \mathcal T_{\pi_a\pi_b\to \pi_c\pi_d} = i D_{\tau\tau}(s) \delta_{ab}\delta_{cd} 
    +i D_{\tau\tau}(t) \delta_{ac}\delta_{bd}  
    +i D_{\tau\tau}(u) \delta_{ad}\delta_{bc} \,,
\label{eq:T}
\end{align}
where $D_{\tau\tau}(p^2)$ is the propagator of the shifted $\chi$ field, $\tau$, such that $\langle \tau \rangle =0$ (for solution I, the VEV $\chi=m_\pi^2$ and for solution II, $\chi=(m_\pi^{II})^2$)~\cite{Chivukula:1991bx,Lyu:2024lzr},
\begin{align}\label{eq:prop of tau}
    D_{\tau\tau}(p^2) &= \frac{i (p^2 -\chi) }{ (p^2 -\chi)  N B(p^2,\chi,M) -v^2}\,,\\
     \label{eq: def of ren-Bfun}
    B(s,\chi,M) &= \frac{1}{32\pi^2} \left( 1 + \rho(s)
\log\frac{\rho(s)-1}{\rho(s)+1}  -\log\frac{\chi}{M^2}\right)\, ,
\end{align}
where $\rho(s) = \sqrt{1-4\chi/s}$.
\begin{figure}
    \centering
    \includegraphics[width=0.4\textwidth]{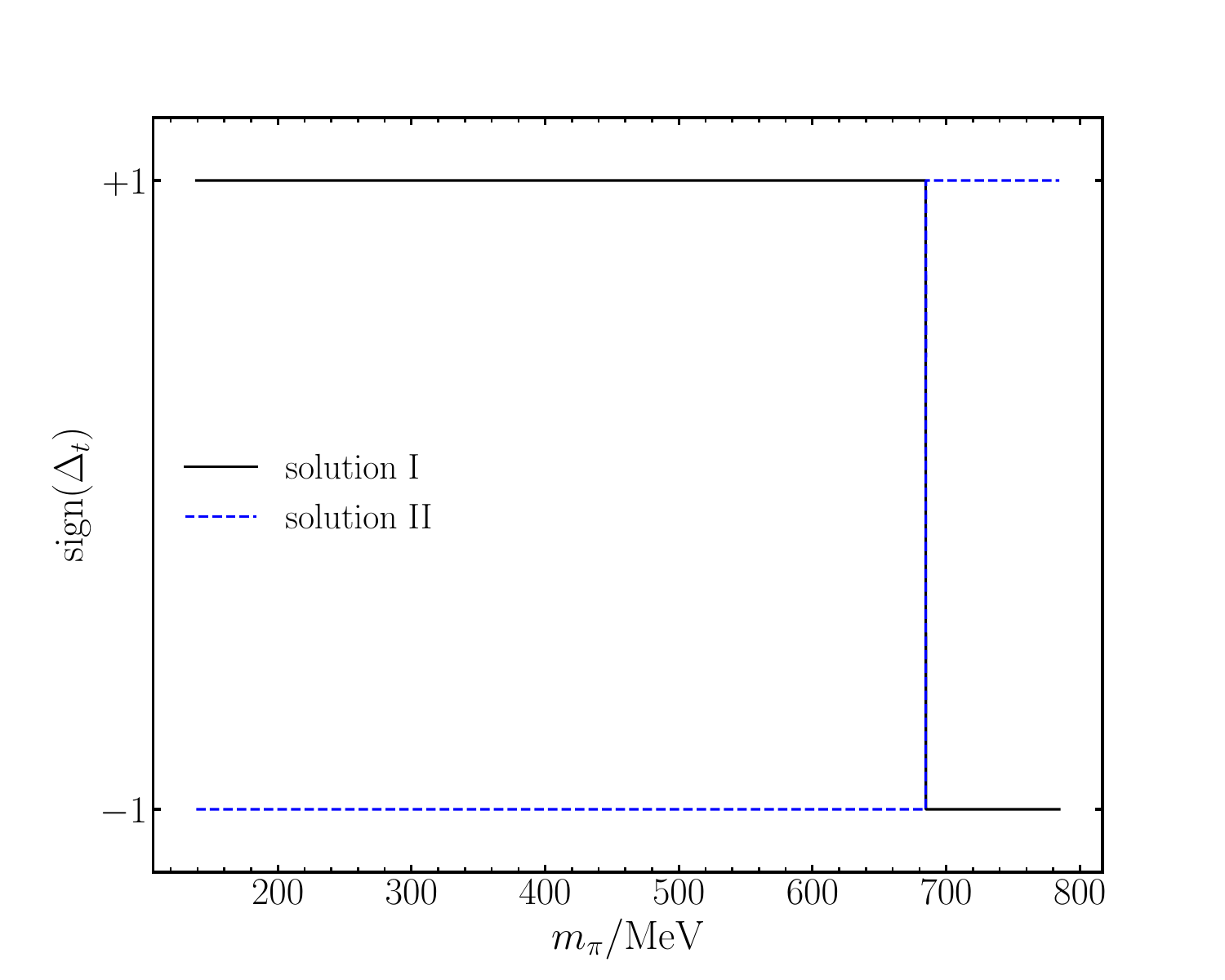}

    \caption{The sign of the criterion function $\Delta_t$ for the existence of a tachyon with regard to the two choices of the vacuum at zero temperature with varying $m_\pi$. Solution I denotes the vacuum at $v=f_\pi$ with $\chi=m_\pi^2$, which is on $V^I(\phi)$ at first and moves to $V^{II}(\phi)$ when $m_\pi>M/\sqrt e$. Solution II is the other vacuum always lying on $V^{II}(\phi)$.}
    \label{fig:tachyon criterion zeroT} 
\end{figure}
The tachyon pole is located on the negative real axis of the $s$-plane with the pole position $s = - m_t^2$, which is identified as the zero point of the inverse of Eq.~\eqref{eq:prop of tau} in the $s<0$ region. Thus the existence of tachyon can be analytically determined by a thorough analysis on the zero point distribution of the function $i D^{-1}_{\tau\tau}(s)$ along the $s<0$ axis. Actually, there is an easy way out: inspired by the analyses of Ref.~\cite{Abbott:1975bn}, in the following we will prove that $i D^{-1}_{\tau\tau}(s)$ is real and monotonically increasing when $s<0$. To simplify the discussion, for $s=-r\,(r>0)$, let $x=\sqrt{1+4\chi/r} \in (1,+\infty)$ and define $f\left(x,\chi,M\right)\equiv B\left(-r,\chi,M\right)$, then
\begin{align}
    f(x,\chi,M) &= \frac{1}{32\pi^2}\left( 1+x\log\frac{x-1}{x+1} -\log\frac{\chi}{M^2}\right)\,, \\
    \frac{\partial f}{\partial x} &=  \frac{1}{32\pi^2}\left( \log\frac{x-1}{x+1} + \frac{2x}{x^2-1} \right) \,.
\end{align} 
It is obvious that $f\to-\infty$ as $r\to\infty$. Furthermore, by defining $z\equiv(x-1)/(x+1) \in (0,1)$ and  $g(z)\equiv 32\pi^2\partial f/\partial x$, we have
\begin{align}
    g(z) &= \log z +\frac{1}{2}\left( \frac{1}{z} - z \right)\,,  \\
    g'(z) &= -\frac{1}{2}\left(\frac{1}{z}-1 \right)^2\,.
\end{align}
Since $g'(z)<0$, then $g(z)>g(1)=0$ with $z \in (0,1)$ and thus $\partial f/\partial x >0$. Considering the definition of the argument $x$, this means that the function $B(s,\chi,M)$ is monotonically increasing along the $s<0$ axis. It is trivial that the rest part of $iD^{-1}_{\tau\tau}(s)$,  $ v^2/(\chi-s)$, is also monotonically increasing in that region. Thus the previous argument has been proved. Additionally, when $s<0$, noting that $iD^{-1}_{\tau\tau}(s) \to -\infty$ as $s\to -\infty$, there can only be no more than one zero point, i.e. tachyon pole. The criterion for the existence of a tachyon is given by requiring the sign of $iD^{-1}_{\tau\tau}(s)$ at $s=0$ to be positive:
\begin{align}\label{eq:tachyon criterion zeroT}
    \Delta_t \equiv \frac{1}{32\pi^2}\left( \log\frac{M^2}{\chi} -1 \right)+\frac{v^2/N}{\chi} >0\,,
\end{align}
which is very similar to the local minimum criterion Eq.~\eqref{eq:criterion of Veff extremum zeroT} and can be verified that they have an identical sign-changing point at $\chi= 32\pi^2f_\pi^2/(Ny_0)$ (with $y_0 e^{y_0} = \frac{32\pi^2 e f_\pi^2}{N M^2} $), corresponding to the situation when solution I and II meet at the same point. Then there will be a tachyon pole if the  criterion Eq.~\eqref{eq:tachyon criterion zeroT} is satisfied and no tachyon pole when the criterion function $\Delta_t$ is negative. The result  for the existence  of a tachyon is shown in Fig.~\ref{fig:tachyon criterion zeroT} which shows the sign of the criterion function $\Delta_t$ in Eq.~\eqref{eq:tachyon criterion zeroT}.  For the local minimum  on $V^I(\phi)$ and the saddle point on $V^{II}(\phi)$ , there is always a tachyon, which results from the fact that there exists another vacuum state on $V^{II}(\phi)$ with lower energy. 

However, the existence of a tachyon for solution I is not a disaster~\cite{Chivukula:1991bx}. If we just regard the $O(N)$ $\sigma$ model as an effective theory of low energy pion dynamics and the validity range of $O(N)$ model can be approximately taken to be $s\gg -m_t^2$ and $s-s_{th} \ll m_t^2$ for real $s$ values, where $s_{th} = 4m_\pi^2$ is the $\pi\pi$ threshold.\footnote{Considering the partial wave projection (e.g., see Eq.~\eqref{eq:ONamp Itu finiteT}), the tachyon poles in the $t$- and $u$-channel will generate an unphysical cut along the region $[s_{th}+m_t^2,+\infty)$, which limits the applicability of $O(N)$ model in the physical region.} In this sense, the tachyon pole indeed serves as a cutoff scale\footnote{Actually, the residue of the tachyon pole in the leading $1/N$ order $\sigma$ propagator is found to be positive,  see Sec.~\ref{sect:Veff_finiteT__subsec:tachyon} for details.}.

\subsection{The particle spectrum for the second branch}

\begin{figure}
    \centering
    \includegraphics[width=0.19\textwidth]{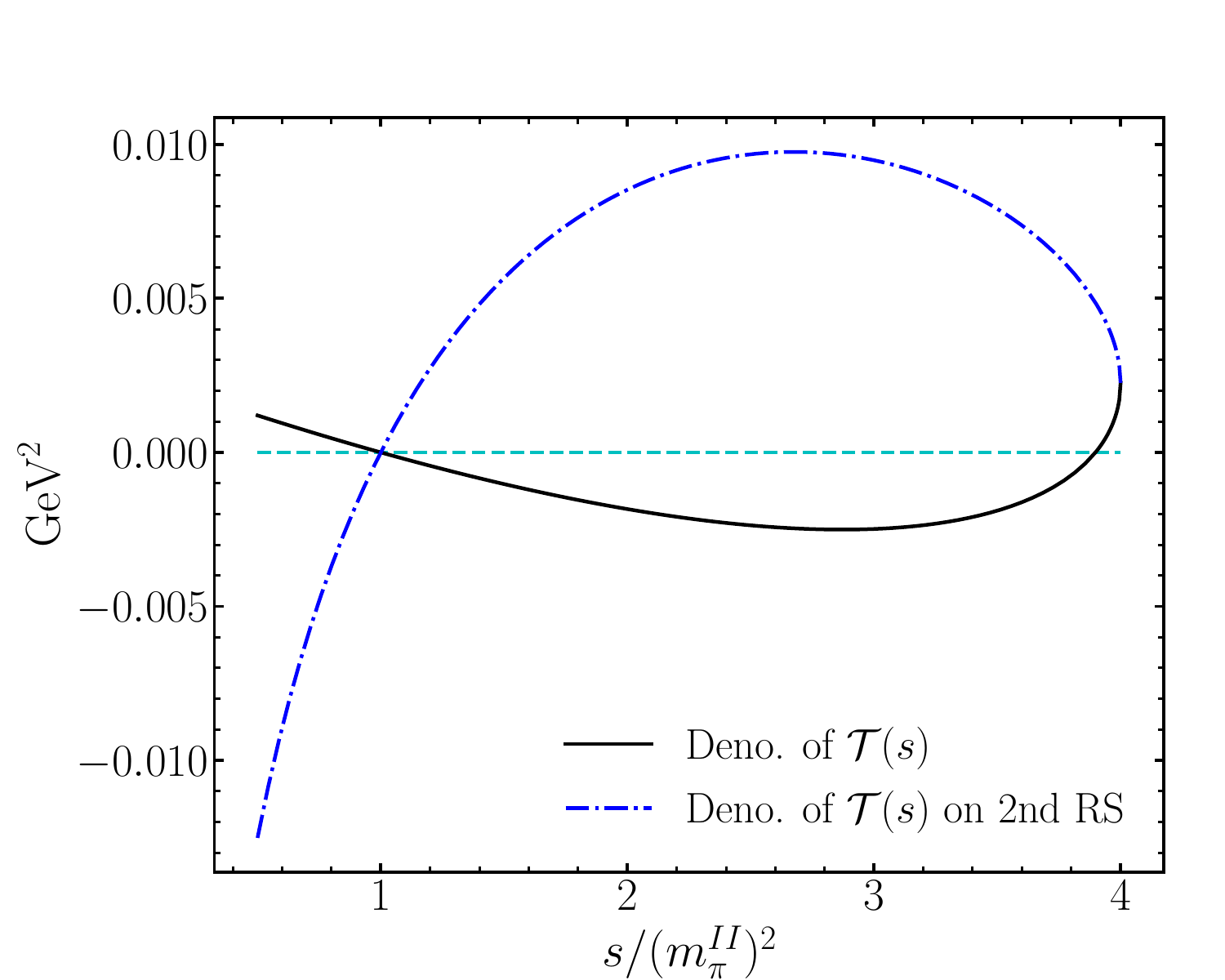}
    \includegraphics[width=0.19\textwidth]{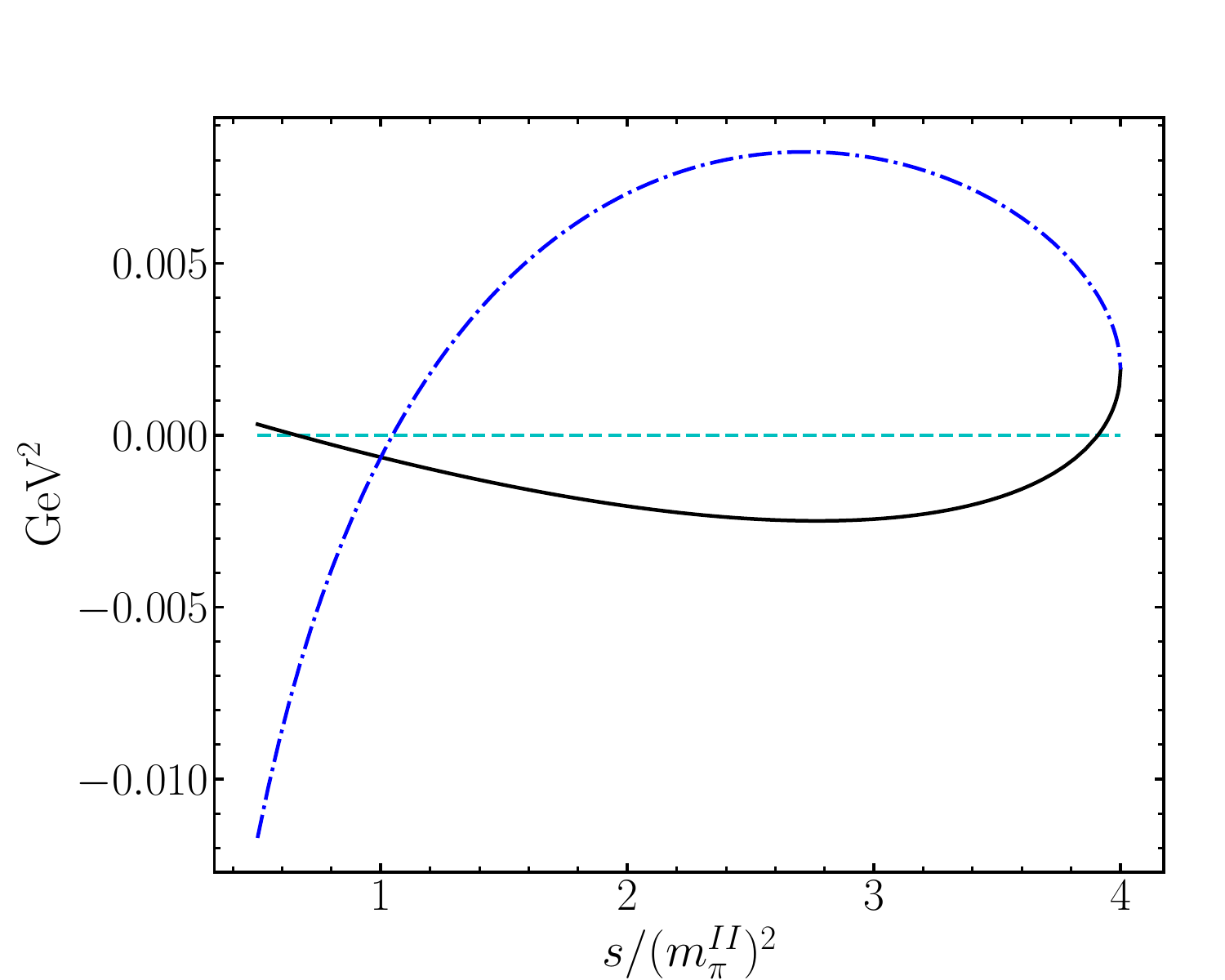}
    \includegraphics[width=0.19\textwidth]{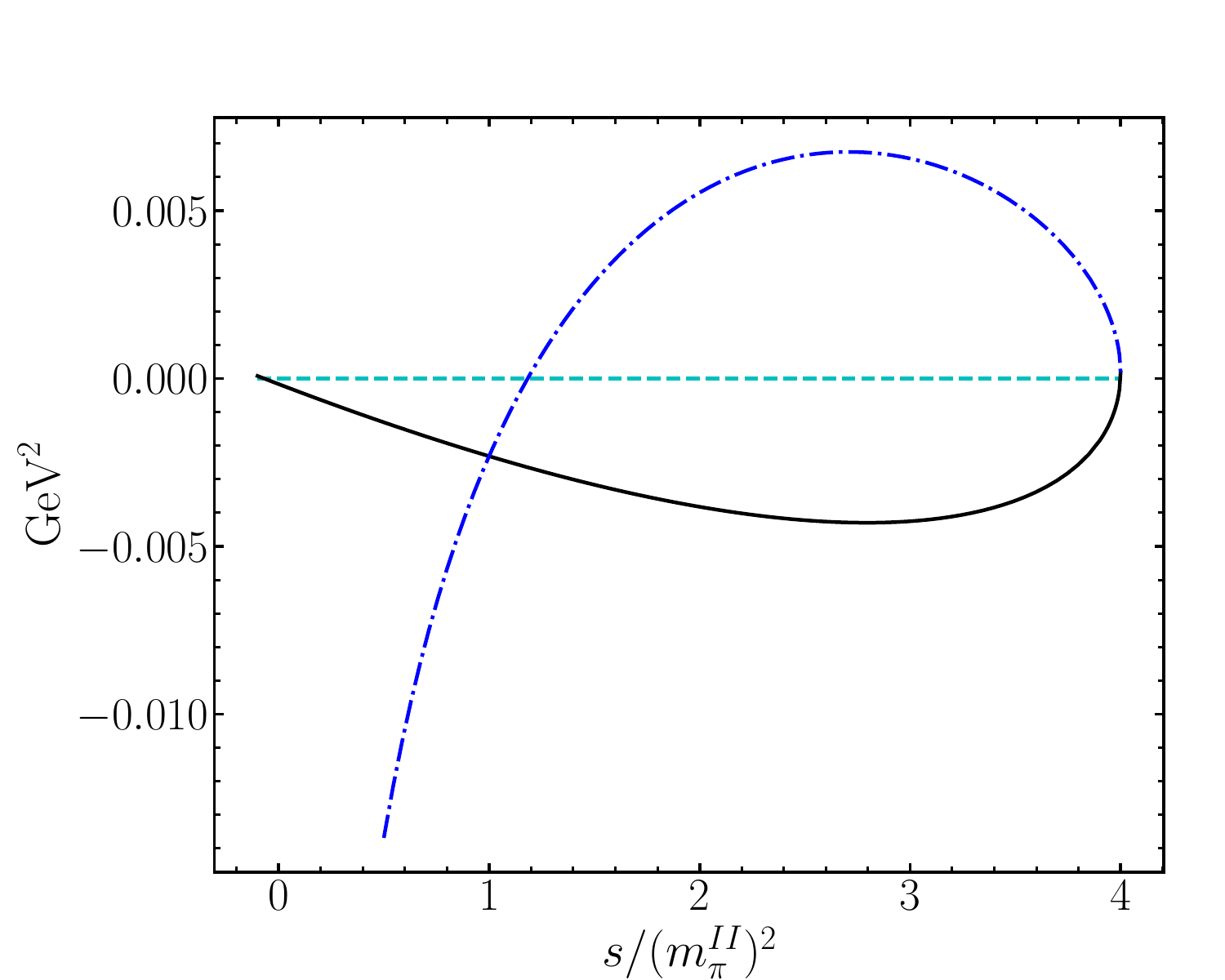}
    \includegraphics[width=0.19\textwidth]{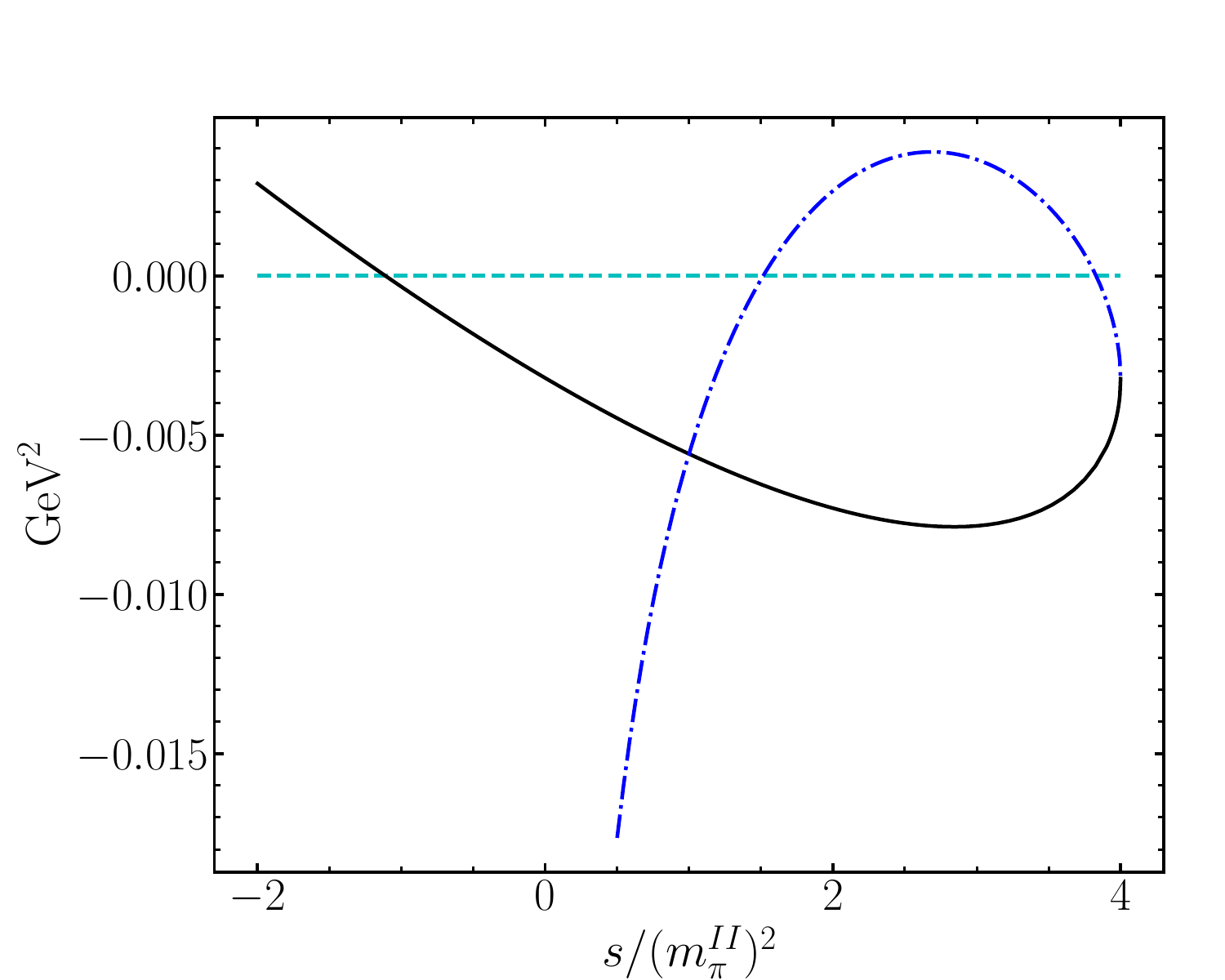}
    \includegraphics[width=0.19\textwidth]{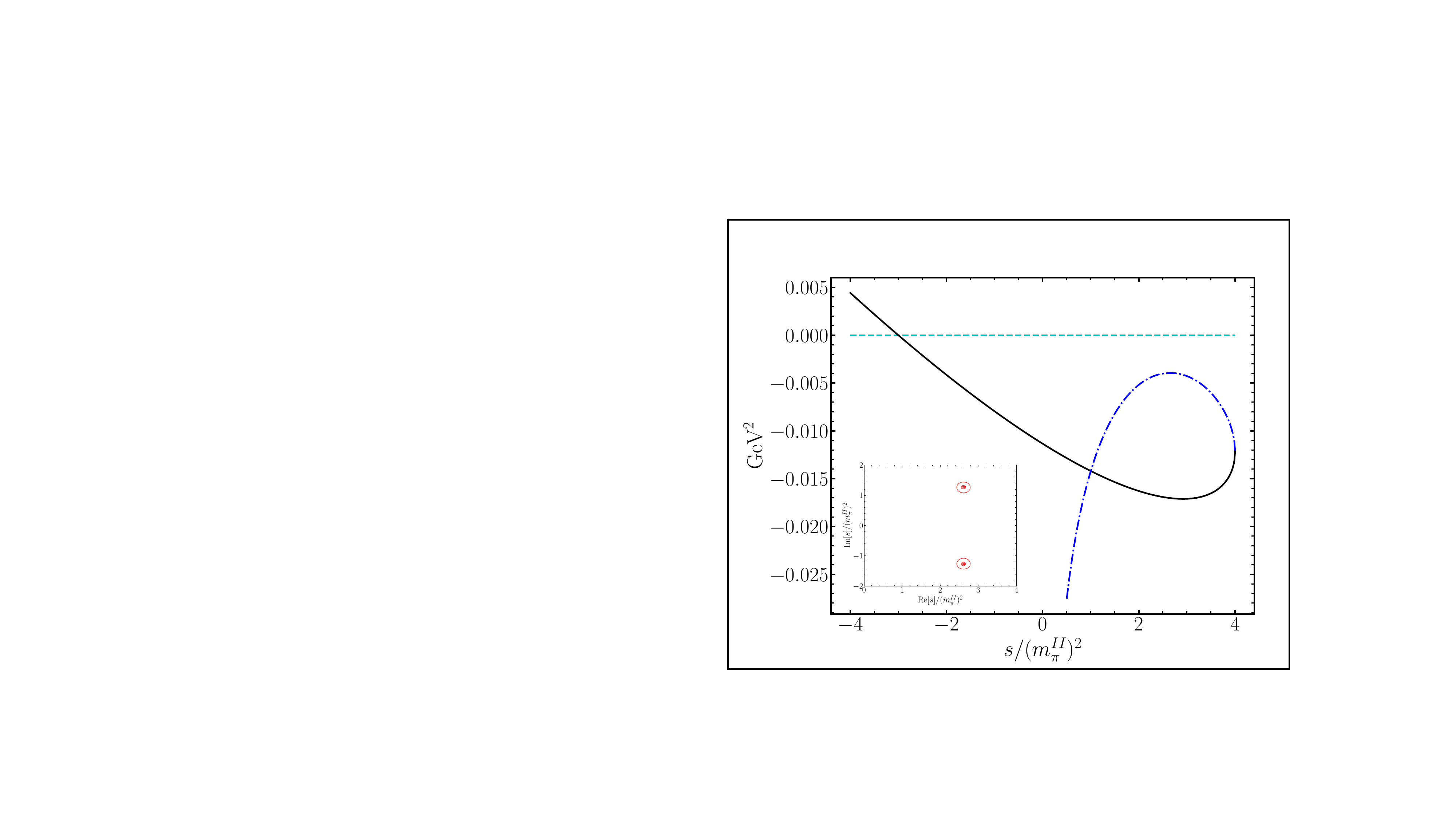}

    \caption{ The subthreshold behaviors of the denominator (Deno.) of the leading $1/N$ order $IJ=00$ channel amplitude $\mathcal T(s)$ on both the physical and second Riemann sheet (RS) for solution II on the second branch. The zero points of the denominator of the amplitude correspond to bound state (or tachyon) poles on first RS and to virtual state poles on second RS. From left to right, the cases are discriminated by various pion masses for solution I of the gap equations, i.e. $m^I_\pi=139$, $500$, $700$, $900$, $1200$ MeV respectively. The inset in the last graph with $m^I_\pi= 1200$ MeV is a contour plot of $|\mathcal T(s)|$ on the second RS, showing the resonance pole in the complex $s$-plane.}
    \label{fig:particle spectrum sol2}  
\end{figure}

Although the solution II on the second branch cannot describe the low energy pion physics, as discussed before, we shall study its particle spectrum with varying $m_\pi$ for completeness. At zero temperature, the $IJ=00$ channel particle spectrum for the vacuum solution II on the second branch of $V(\phi)$ can be extracted from the partial wave amplitude~\cite{Coleman:1974jh,Chivukula:1991bx,Lyu:2024lzr},
\begin{align}
    \mathcal T^{LO}_{00}(s)=\frac{i N D_{\tau \tau}(s)}{32\pi} = -\frac{1}{32\pi} \frac{s -(m_\pi^{II})^2 }{ (s -(m_\pi^{II})^2)
    B(s,(m_\pi^{II})^2,M) -(f_\pi^{II})^2/N}\,,
\label{eq:LO-T00}
\end{align}
where the $m_\pi$ and $f_\pi$ values are set for solution II  of the gap equations Eqs.~(\ref{eq:zeroT Gap eq 1}, \ref{eq:zeroT Gap eq 2}), and $B(s,(m_\pi^{II})^2,M)$ is defined in Eq.~\eqref{eq: def of ren-Bfun}.  The results with different pion mass values of solution I ($m_\pi^I$) are shown in Fig.~\ref{fig:particle spectrum sol2}.
When $m_\pi^I$ increases from its physical value $139$ MeV, the $IJ=00$ particle spectrum for solution II, which always lies on the second branch of the effective potential, demonstrates four different types of contents sequentially: (i) two bound states and one virtual state, (ii) one bound state, one virtual state and one tachyon, (iii) two virtual states and one tachyon, and (iv) one resonance and one tachyon. The criterion \eqref{eq:tachyon
criterion zeroT} also applies for the (ii) and (iii) cases. Moreover, for $m_\pi^I$ not too far from the physical value (e.g., $m_\pi^I<\sqrt{\chi_b}$),  the value of $m_\pi^{II}$ is always much larger than that of solution I and hence results in a much smaller $f_\pi^{II}$ value, which can be found out in Fig.~\ref{fig:vT_mpiT_varying_mpi_zeroT} and also in Eqs.~(\ref{eq:zeroT Gap eq 1}, \ref{eq:zeroT Gap eq 2}). The physics described by solution II is very different compared to the particle spectrum and pole trajectories shown in Ref.~\cite{Lyu:2024lzr}, where the vacuum is actually always chosen to be solution I, originally located on the first branch of the effective potential. Thus in the $O(N)$ model, for pion masses not much larger than the physical value, the phenomenologically preferred vacuum is on the first branch, not the second branch.

\section{The vacuum structure of $O(N)$ $\sigma$ model with varying $m_\pi$ at finite temperature\label{sect:Veff_finiteT}}

The finite temperature effective potential of $O(N)$ model with explicit symmetry breaking at leading order $1/N$ expansion is~\cite{Andersen:2004ae}:
\begin{align}
    V^{T}(\phi,\chi) &= V(\phi,\chi) + N I^{T\neq 0}(\chi)  \label{eq:Veff finite T}\,,\\
    I^{T\neq 0}(\chi) &= -\frac{1}{6\pi^2}\int^\infty_0 \mathrm d k \frac{k^4}{\omega_k(\chi)}n_B(\omega_k(\chi))\,,
\end{align}
where $\beta =1/T$, $\omega_k(\chi) = \sqrt{k^2 + \chi}$ and $n_B(\omega_k) = (e^{\beta\omega_k}-1)^{-1}$ is the Bose-Einstein distribution.
The finite temperature contribution $I^{T\neq 0}$ can be standardly obtained using the imaginary time formalism.
\begin{figure}%[!htbp]       
    \centering
    \includegraphics[width=0.4\textwidth]{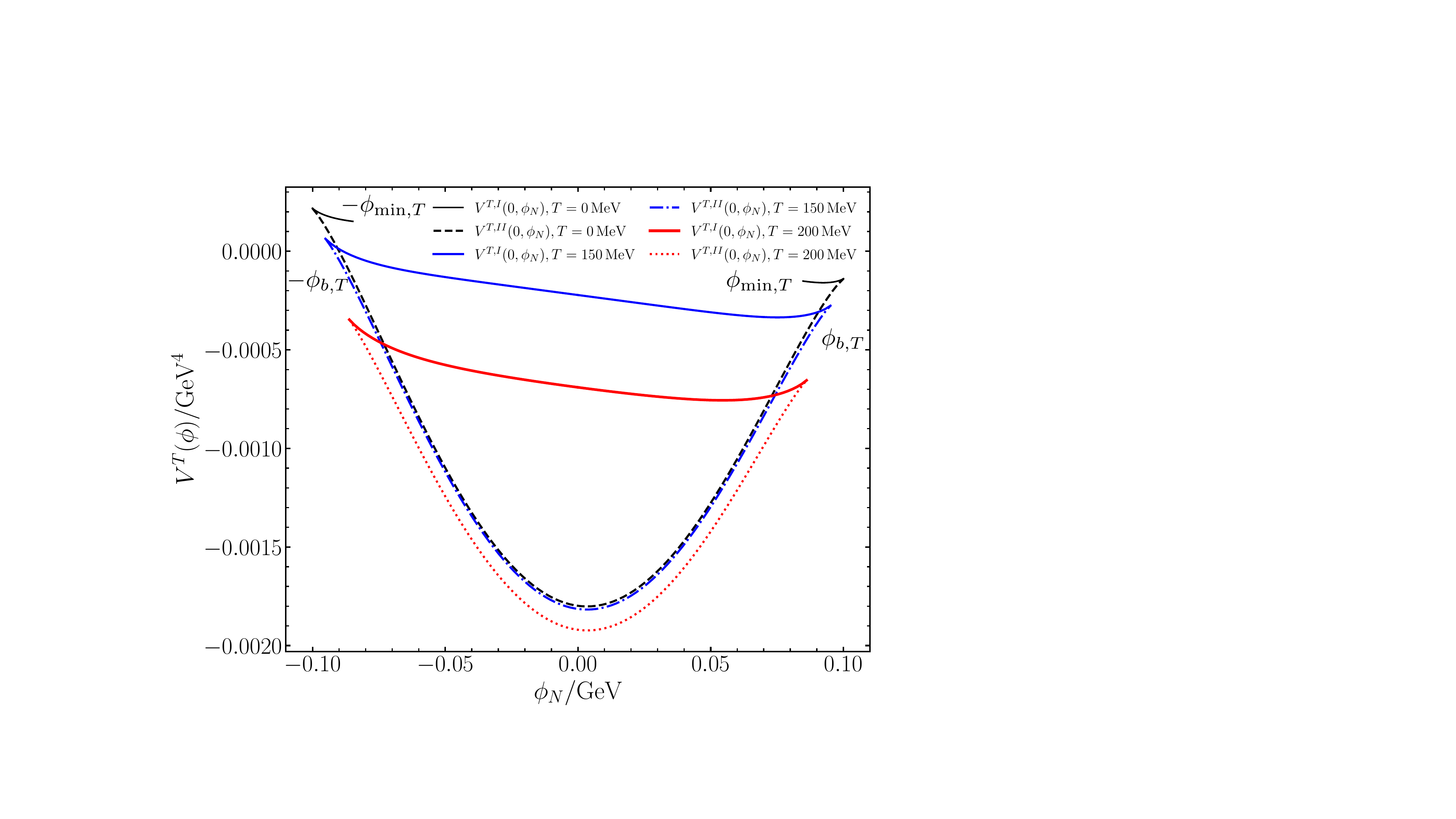}
    \includegraphics[width=0.4\textwidth]{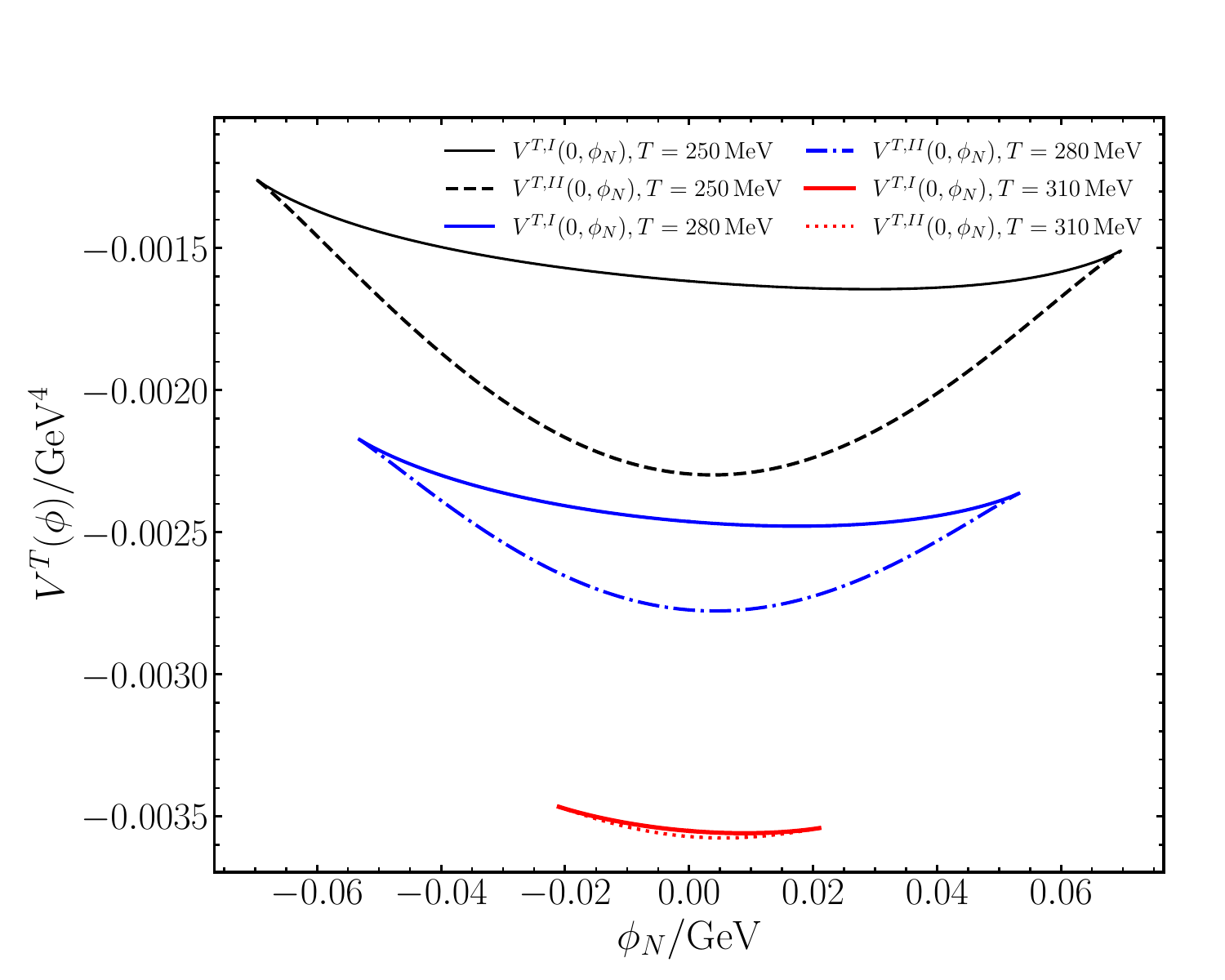}
    \caption{Effective potential at finite temperature. For clearness of the presentation, the plots are limited to a ``slice'' of the effective potential along the $\phi_N$ direction with $\phi_a(a<N)$ set to their vacuum expectation values $\phi_a =0(a<N)$. The plots only show the region where the effective potential is real.  The cases correspond to different temperature values $T= 0$, $150$, $200$, $250$, $280$ and $310$ MeV respectively. The zero-temperature values are set as $v^I(0)=f_\pi$ and $m_\pi^I(0) = 139\,\mathrm{MeV}$.} 
    \label{fig:Veff finiteT}
\end{figure}
The gap equations are then obtained again by the requirements $\partial V /\partial \chi =0$ and $\partial V/ \partial \phi_a =0$,
\begin{align}\label{eq:finiteT Gap eq 1}
    \phi^2 &= f_\pi^2 +\frac{N}{16\pi^2}\left( m_\pi^2 \log\frac{m_\pi^2}{M^2} - \chi  \log\frac{\chi }{M^2}  \right) - N A^{T\neq 0} (\chi)\,, \\
    \chi \phi_a &= 0\,(a<N)\,,\quad 
    \chi \phi_N = \alpha\,,
    \label{eq:finiteT Gap eq 2}
\end{align}
where $\phi^2 = v^2(T)$ and $\chi = m^2_\pi(T)$. The function $A^{T\neq 0}$ is defined as the finite temperature contribution to the tadpole integral encountered in the derivation of Eq.~\eqref{eq:zeroT Gap eq 1},
\begin{align}
    A^{T\neq 0} (\chi) = \int^\infty_0 \frac{\mathrm d k}{2\pi^2} \frac{k^2 n_B(\omega_k(\chi))}{\omega_k(\chi)}\,.
\end{align}
The function $\phi^2(\chi)$ defined by Eq.~\eqref{eq:finiteT Gap eq 1} has the same typical behavior as the one at $T=0$ in Fig.~\ref{fig:phi_sq of chi} (with $\chi_b$, $\phi^2_b$ and $\phi^2_\text{min}$ varying with temperature, i.e. $\chi_{b,T}$, $\phi^2_{b,T}$ and $\phi^2_{\text{min},T}$). 
The effective potential at finite temperature (shown in Fig.~\ref{fig:Veff finiteT}) is obtained by substituting $\chi$ with $\chi^{T,I}(\phi^2)$ and $\chi^{T,II}(\phi^2)$ solved from Eq.~\eqref{eq:finiteT Gap eq 1} respectively ($\chi^{T,I}(\phi^2)\le \chi_{b,T} \le \chi^{T,II}(\phi^2)$ ). Though $V^T(\phi)$ is still double-valued when $\phi^2<\phi_{b,T}^2$, it has different behaviors compared to the case with varying $m_\pi$ in Fig.~\ref{fig:Veff varying mpi zeroT}. As temperature increases, $\phi^2_{\text{min},T}$ and $\phi^2_{b,T}$ decrease. When the former reaches the point $\phi^2_{\text{min},T} = 0$, $V^{T,I}(\phi)$ will be totally real in the region $\phi^2 < \phi^2_{b,T}$; while as the latter moves towards zero, $V^{T,I}(\phi)$ and $V^{T,II}(\phi)$ get closer to each other and at some temperature $T_b$, $\phi_{b,T_b}=0$. In the high temperature regime with $T\gg T_c$, where $T_c= \sqrt{12/N} f_\pi \simeq 160 \,\mathrm{MeV}$ is the critical temperature for chiral restoration in $O(N)$ model without explicit symmetry breaking~\cite{Bochkarev:1995gi,Andersen:2004ae}, there exists another upper bound for the temperature, namely $T_f$, where the gap equations have no solution. It is found numerically and can be analytically verified that $T_f<T_b$,
which means that the system would already be unstable after the
temperature is raised to $T_f$ before it reaches
$T_b$. 

\subsection{The local stability of the effective potential}

\begin{figure}%[!htbp]       
    \centering
    \includegraphics[width=0.3\textwidth]{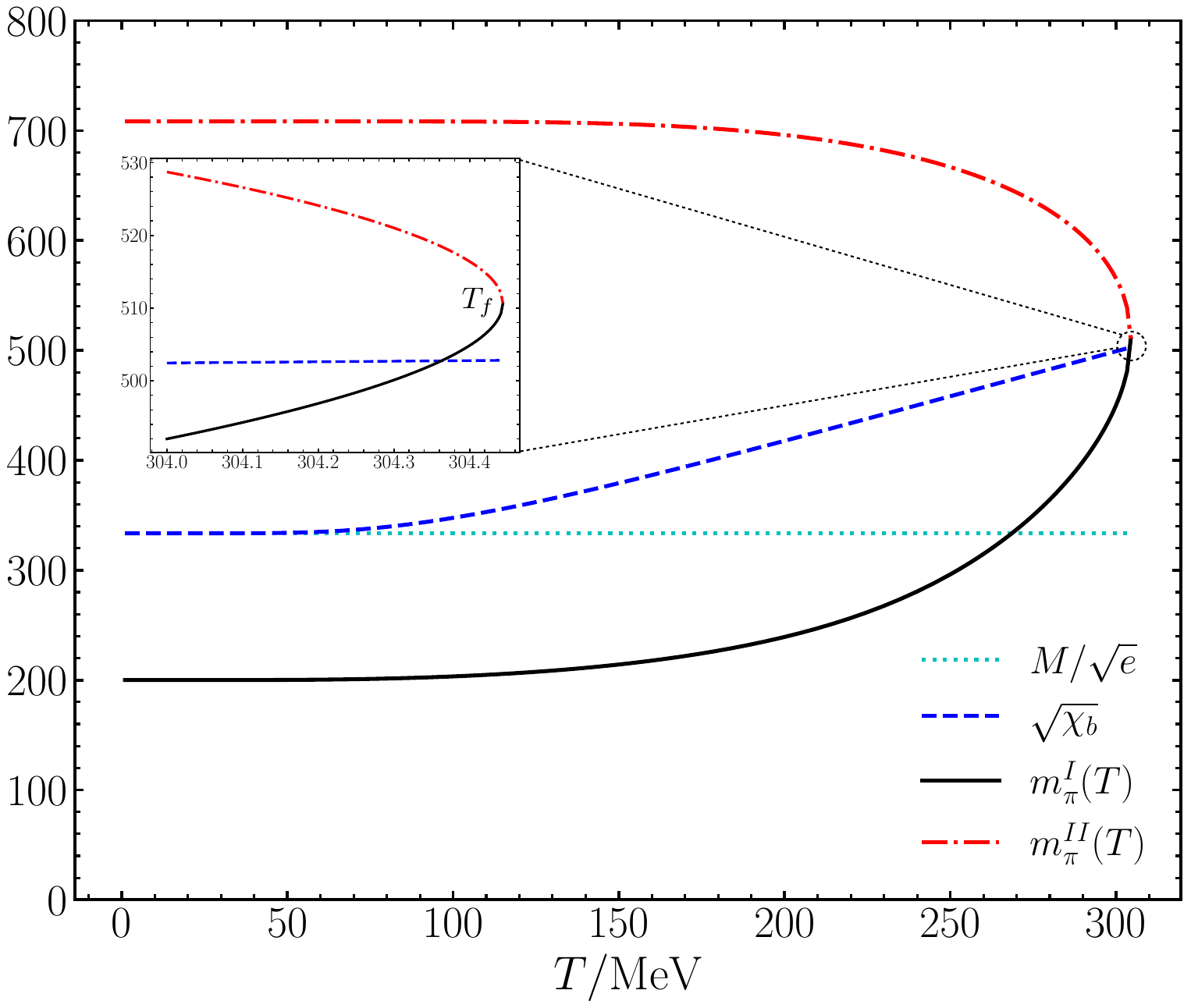}
    \quad
    \includegraphics[width=0.3\textwidth]{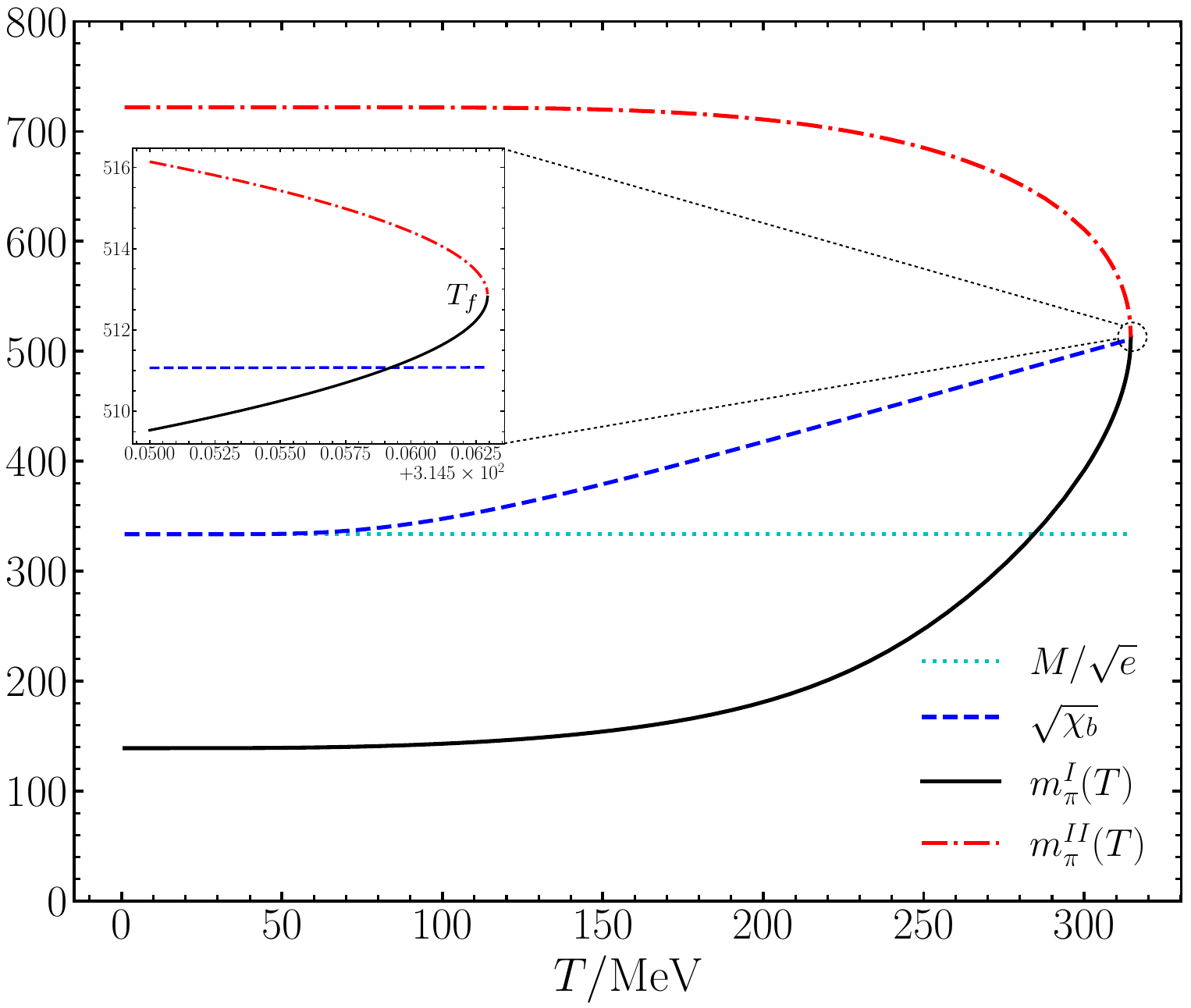}
    \quad
    \includegraphics[width=0.3\textwidth]{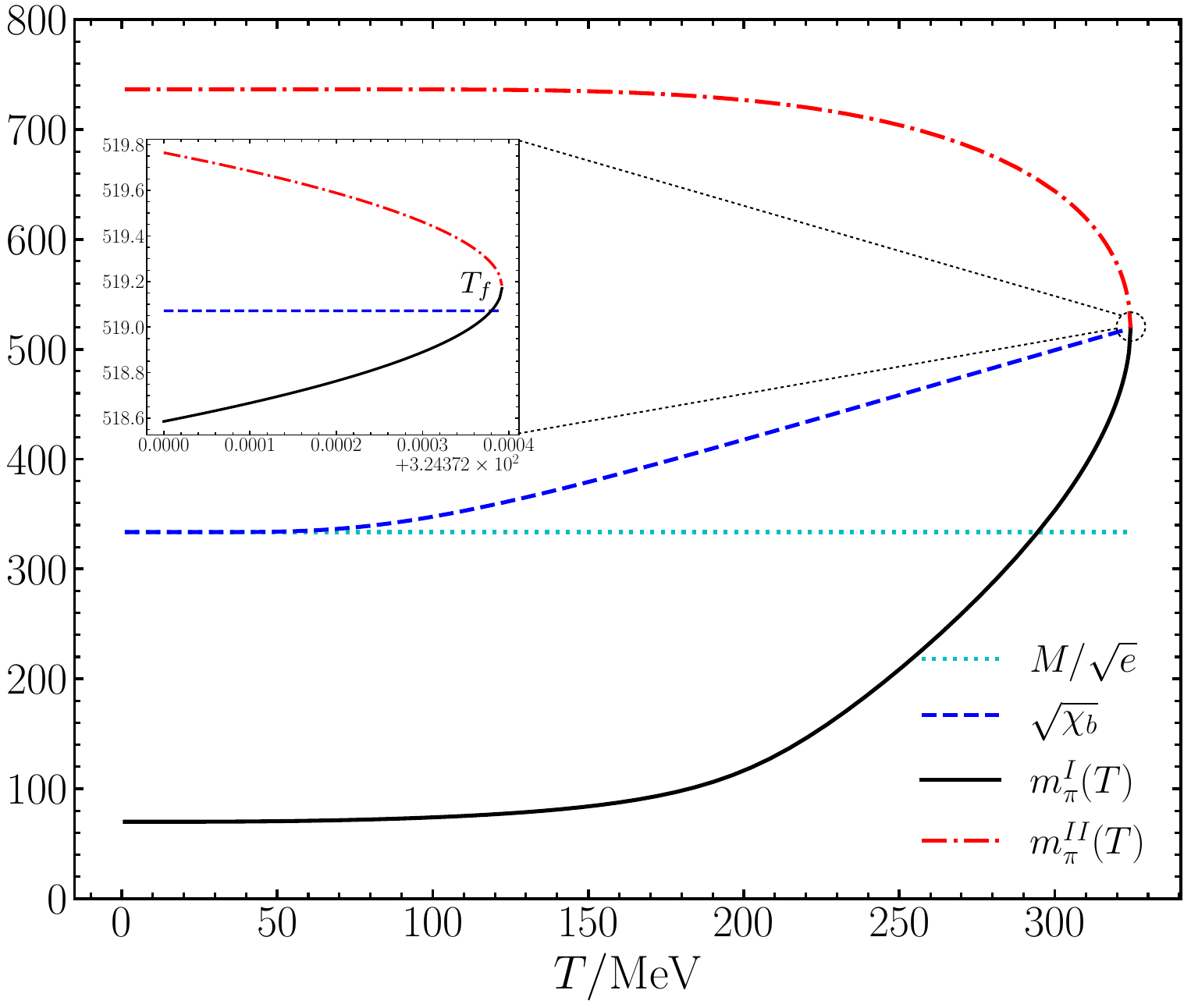}

    \caption{$\sqrt{\chi_{b,T}}$, $m_\pi^{I}(T)$ and $m^{II}_\pi(T)$. From left to right, the three cases are corresponding to zero-temperature values $m_\pi^I(0) = 200$, $139$ and $70$ MeV respectively with $v^I(0) = f_\pi$.}  
    \label{fig:Veff two sols and chi_b} 
\end{figure}

\begin{figure}%[!htbp]       
    \centering
    \includegraphics[width=0.3\textwidth]{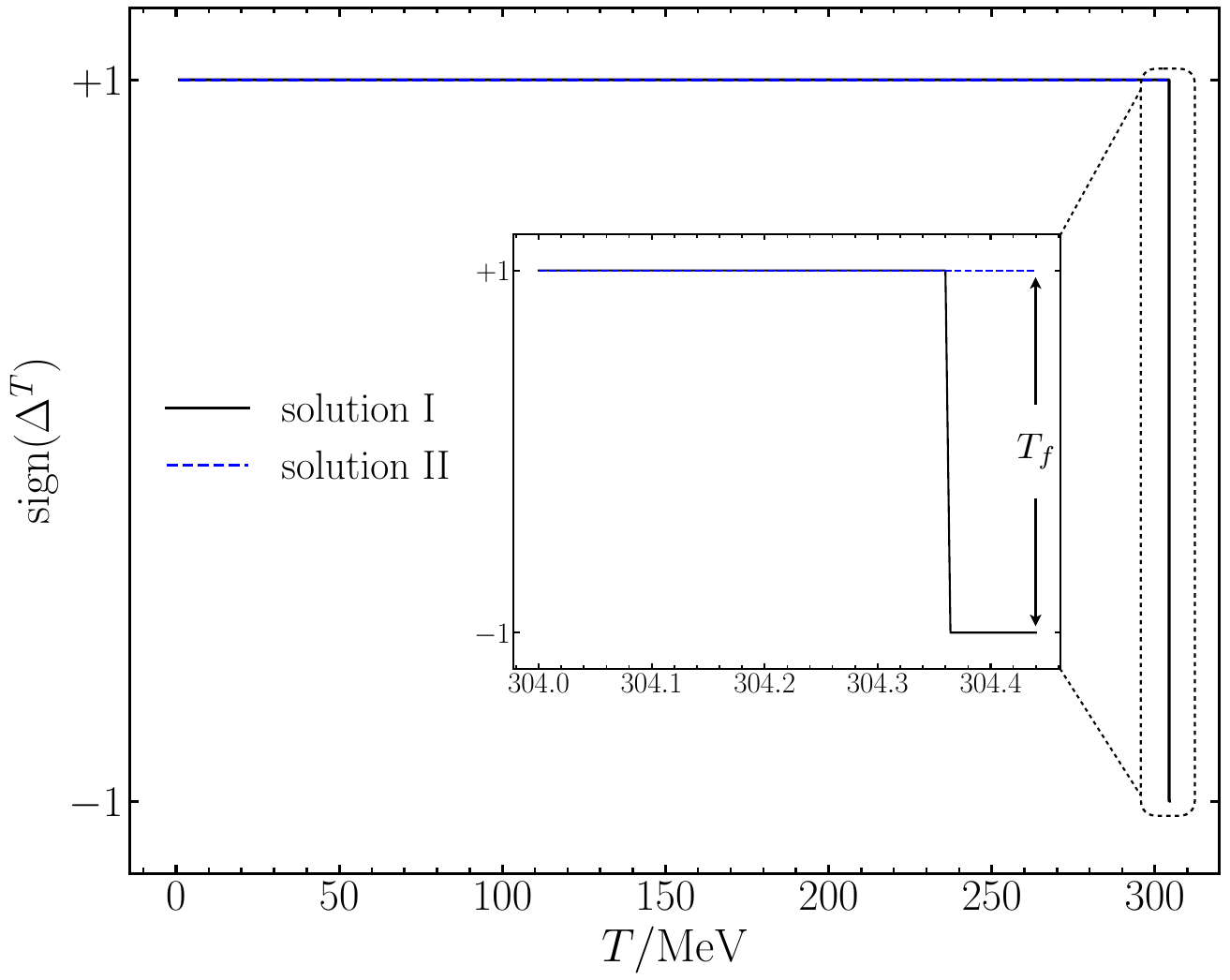}
    \quad
    \includegraphics[width=0.3\textwidth]{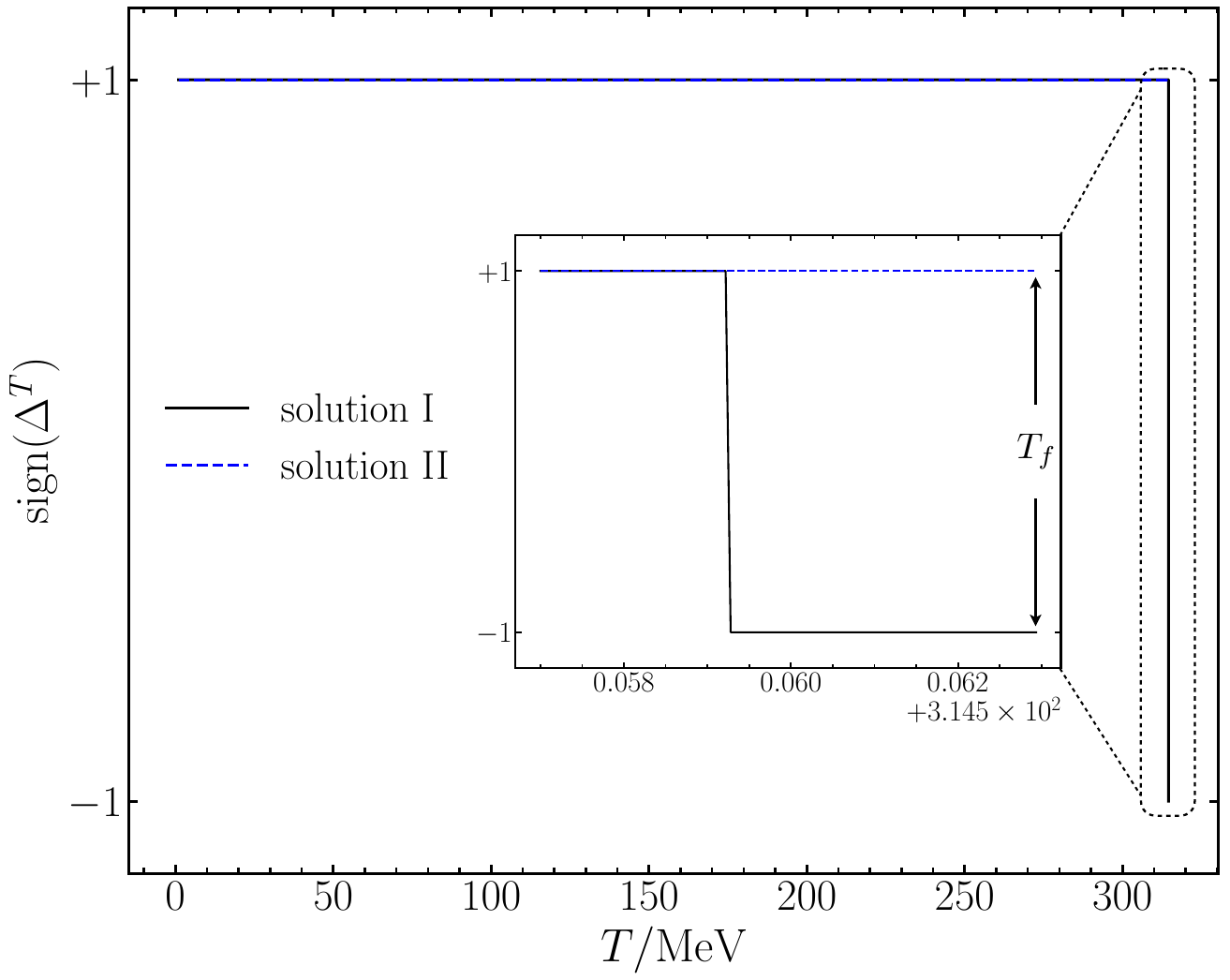}
    \quad
    \includegraphics[width=0.3\textwidth]{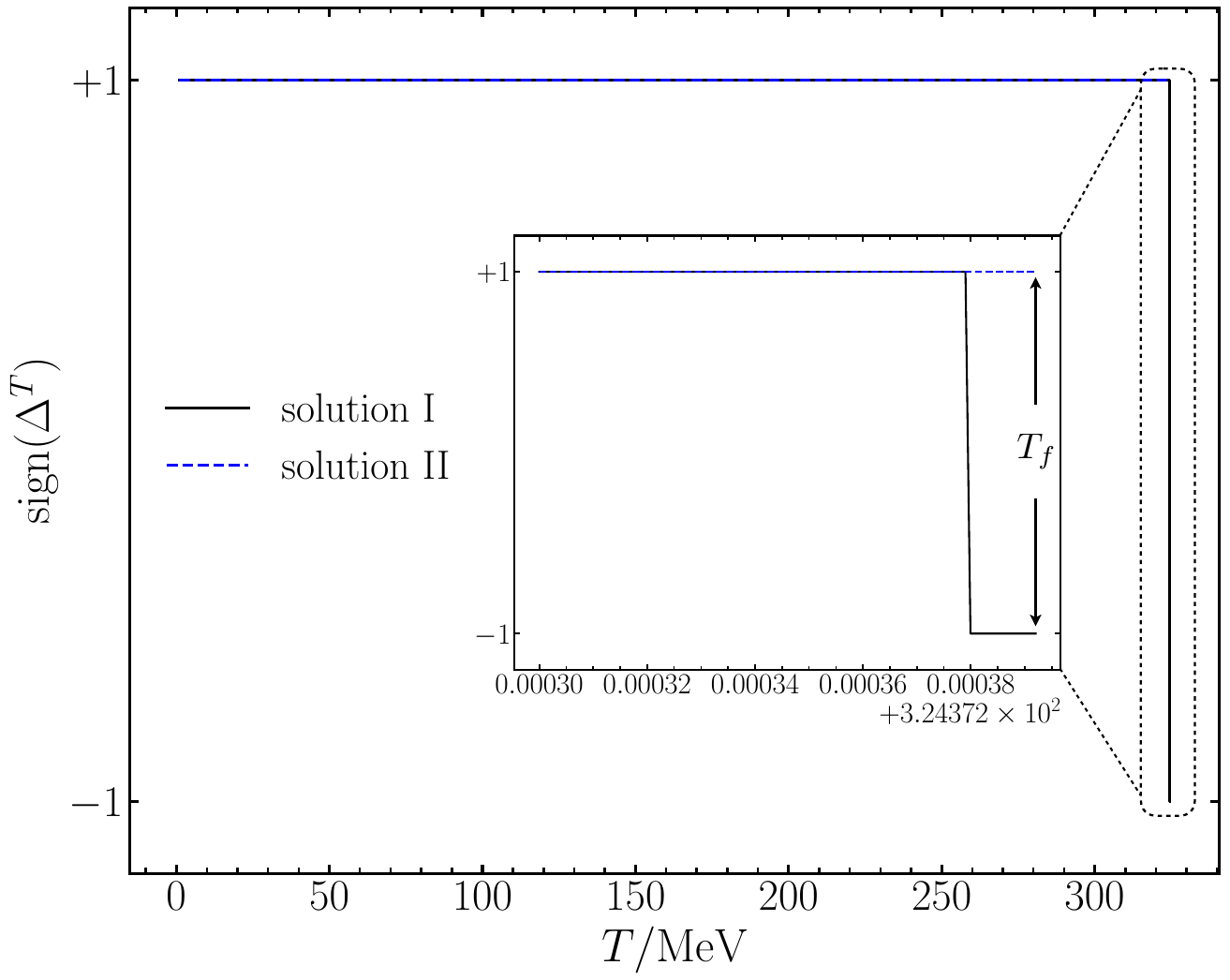}

    \caption{The sign of the criterion function $\Delta^T$ for the solution of the gap equations being a local minimum of effective potential. Solution I denotes the vacuum solution with $v(0)=f_\pi$ and $m_\pi(0)=m_\pi$, which is on $V^{T,I}(\phi)$ at first and moves to $V^{T,II}(\phi)$ when $T\sim T_f$. Solution II is the other vacuum always lying on $V^{T,II}(\phi)$. From left to right, the three cases correspond to zero-temperature values $m_\pi^I(0) = 200$, $139$ and $70$ MeV respectively with $v^I(0) = f_\pi$.}   
    \label{fig:Veff criterion minimum finiteT}  
\end{figure}

At finite temperature, whether the vacuum solutions of the gap equations are indeed local minima of the effective potential remains to be verified. Due to the finite temperature corrections from the loop integrals, the criterion Eq.~\eqref{eq:criterion of Veff extremum zeroT} for the vacuum solution to be a local minimum should be modified as
\begin{align}\label{eq:criterion of Veff extremum finiteT}
    \Delta^T \equiv \frac{32\pi^2v^2(T)/N}{\log(M^2/\chi) -1- 16\pi^2\mathrm d A^{T\neq 0}(\chi)/\mathrm d \chi} +\chi >0\,,
\end{align}
where
\begin{align}
    - 16\pi^2 \frac{\mathrm d}{\mathrm d \chi} A^{T\neq 0}(\chi) = 4\int^\infty_0 \mathrm d k \frac{n_B(\omega_k(\chi))}{\omega_k(\chi)}\,.
\end{align}
With increasing temperature, both $m^I_\pi(T)$ (solution I) and $m_\pi^{II}(T)$ (solution II)
are getting closer gradually to each other and finally meet around $\sqrt{\chi_{b,T}}$ on the second branch of the effective potential at $T=T_f$ (see Fig.~\ref{fig:Veff two sols and chi_b}). For $T>T_f$, as mentioned before, there are no solutions of the gap equations.
With the criterion Eq.~\eqref{eq:criterion of Veff extremum finiteT}, it is found (results shown in Fig.~\ref{fig:Veff criterion minimum finiteT}) that solution I remains a local minimum when it is on $V^I(\phi)$ and becomes a saddle point after moving to the second branch as $T\sim T_f$; whereas solution II is always a local minimum of $V^{II}(\phi)$.

\subsection{The existence of a tachyon\label{sect:Veff_finiteT__subsec:tachyon}}

At finite temperature, the existence of a tachyon can be similarly determined by analyzing the zero point of the inverse of the $\pi\pi$ scattering thermal amplitude in $O(N)$ model (e.g., for $IJ=00$ channel, $\mathcal T^T_{00} = (N/(32\pi)) i D^T_{\tau\tau} $, which has been used in Ref.~\cite{Lyu:2024lzr} to trace the $\sigma$ pole trajectory with varying temperature). With regard to the tachyon pole, it would be enough to study the zero point of $i (D^{T}_{\tau\tau})^{-1}$, 
\begin{align}\label{eq:inverse of Tamp fniteT}
     i (D^{T}_{\tau\tau})^{-1}(s) = N B^{T}(s,\chi,M) - \frac{v^2(T)}{s-\chi}\,,
\end{align}
\begin{figure}%[!htbp]       
    \centering
    \includegraphics[width=0.32\textwidth]{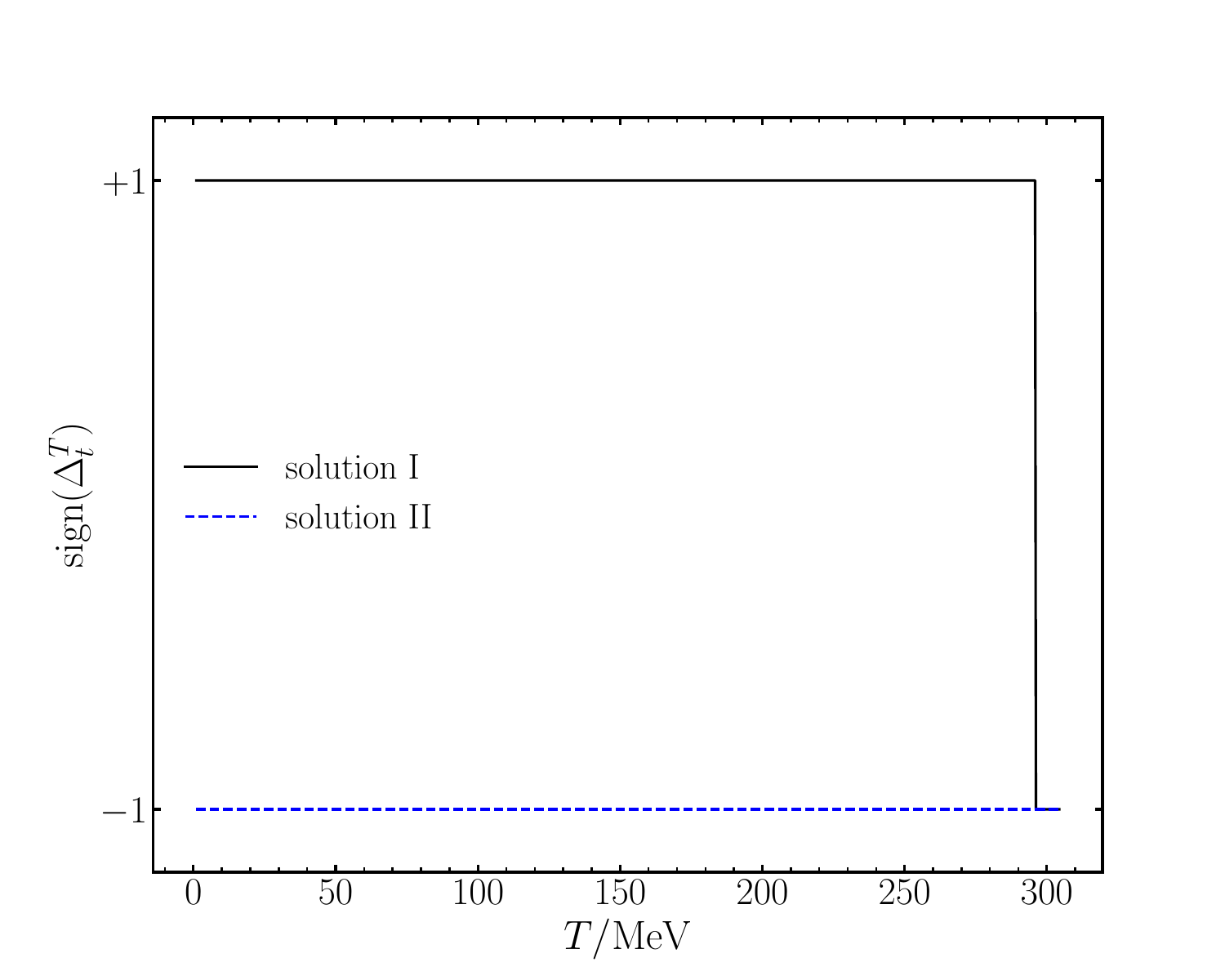}
    \includegraphics[width=0.32\textwidth]{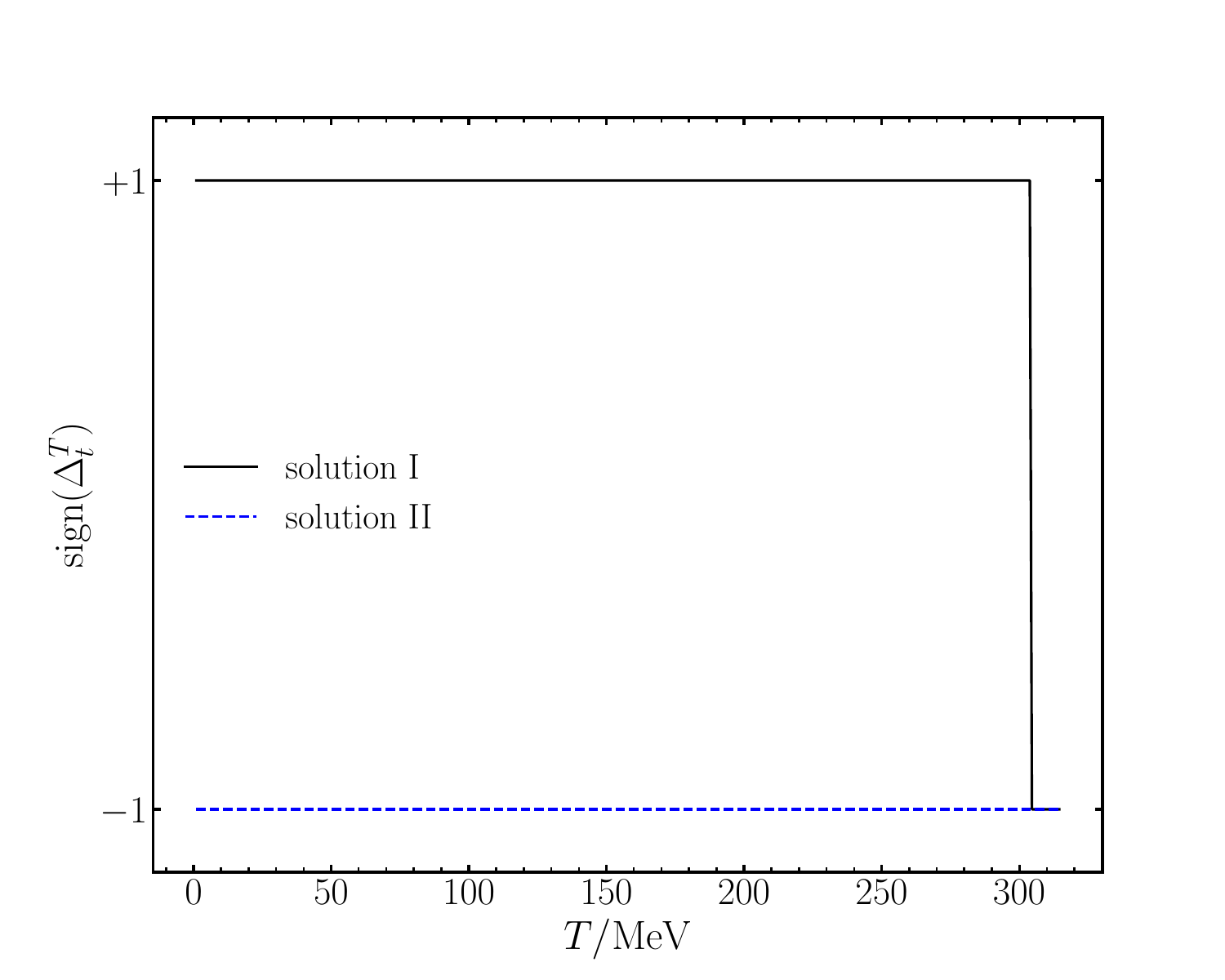}
    \includegraphics[width=0.32\textwidth]{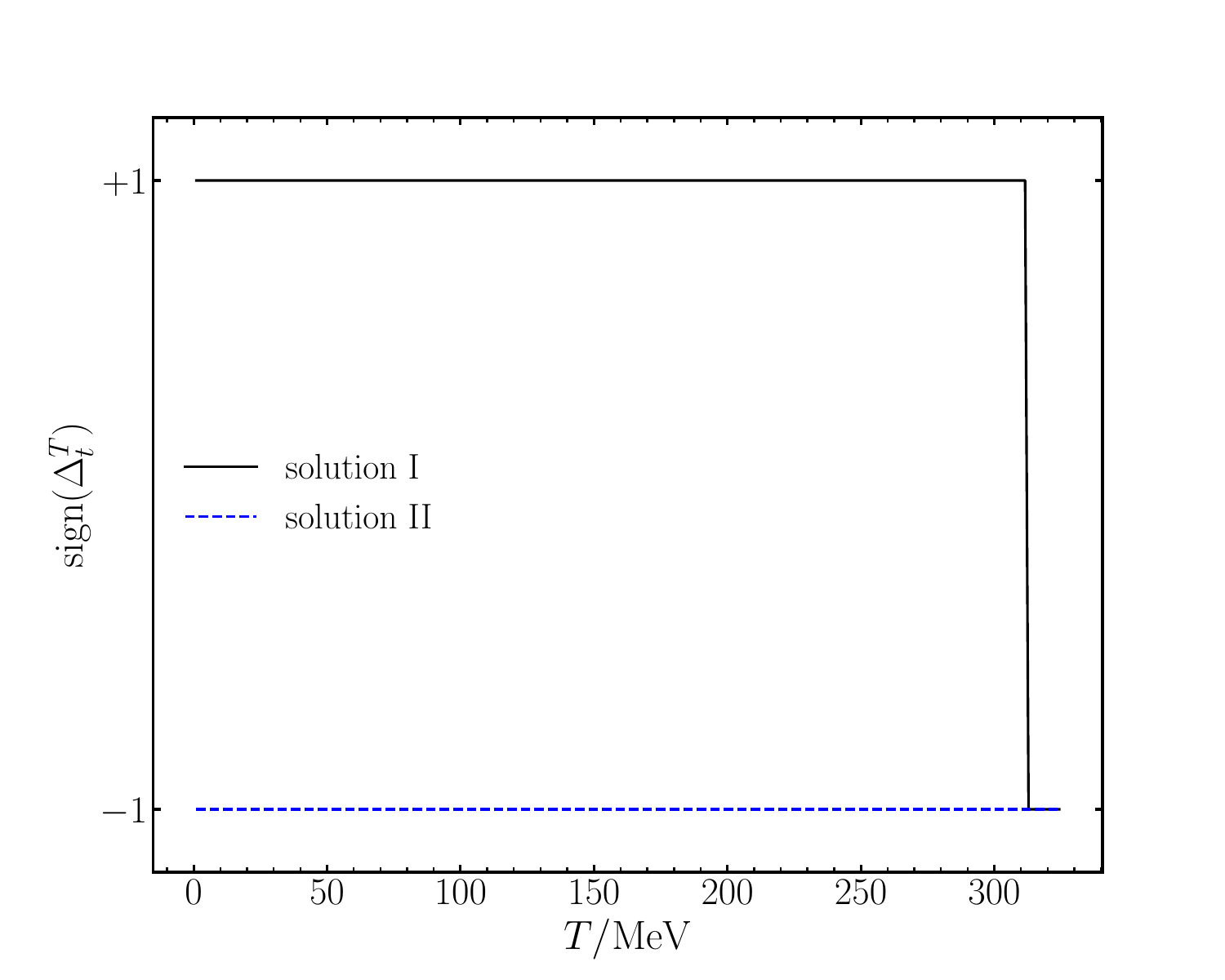}

    \caption{The sign of the criterion function $\Delta^T_t$ for the existence of a tachyon with regard to the two choices for the vacuum. Solution I denotes the vacuum solution with $v(0)=f_\pi$ and $m_\pi(0)=m_\pi$, which is  on $V^{T,I}(\phi)$ at first and moves to $V^{T,II}(\phi)$ when $T\sim T_f$. Solution II is the other vacuum always lying on $V^{T,II}(\phi)$. From left to right, the three cases correspond to $m_\pi^{I}(0) = 200$, $139$ and $70$ MeV respectively with $v^{I}(0) = f_\pi$.}   
    \label{fig:tachyon criterion finiteT}   
\end{figure}
\begin{figure}      
    \centering
    \includegraphics[width=0.4\textwidth]{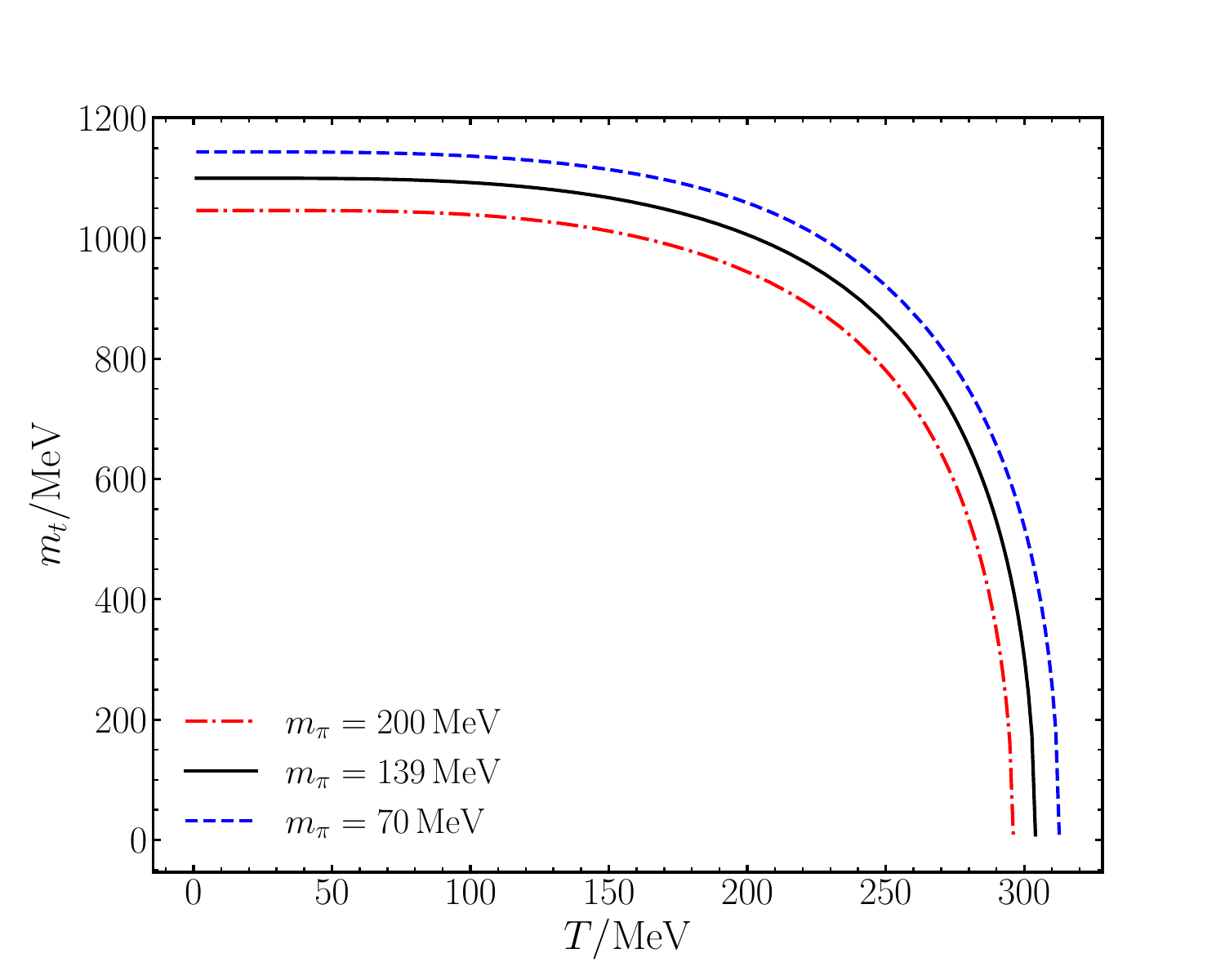}
    \includegraphics[width=0.4\textwidth]{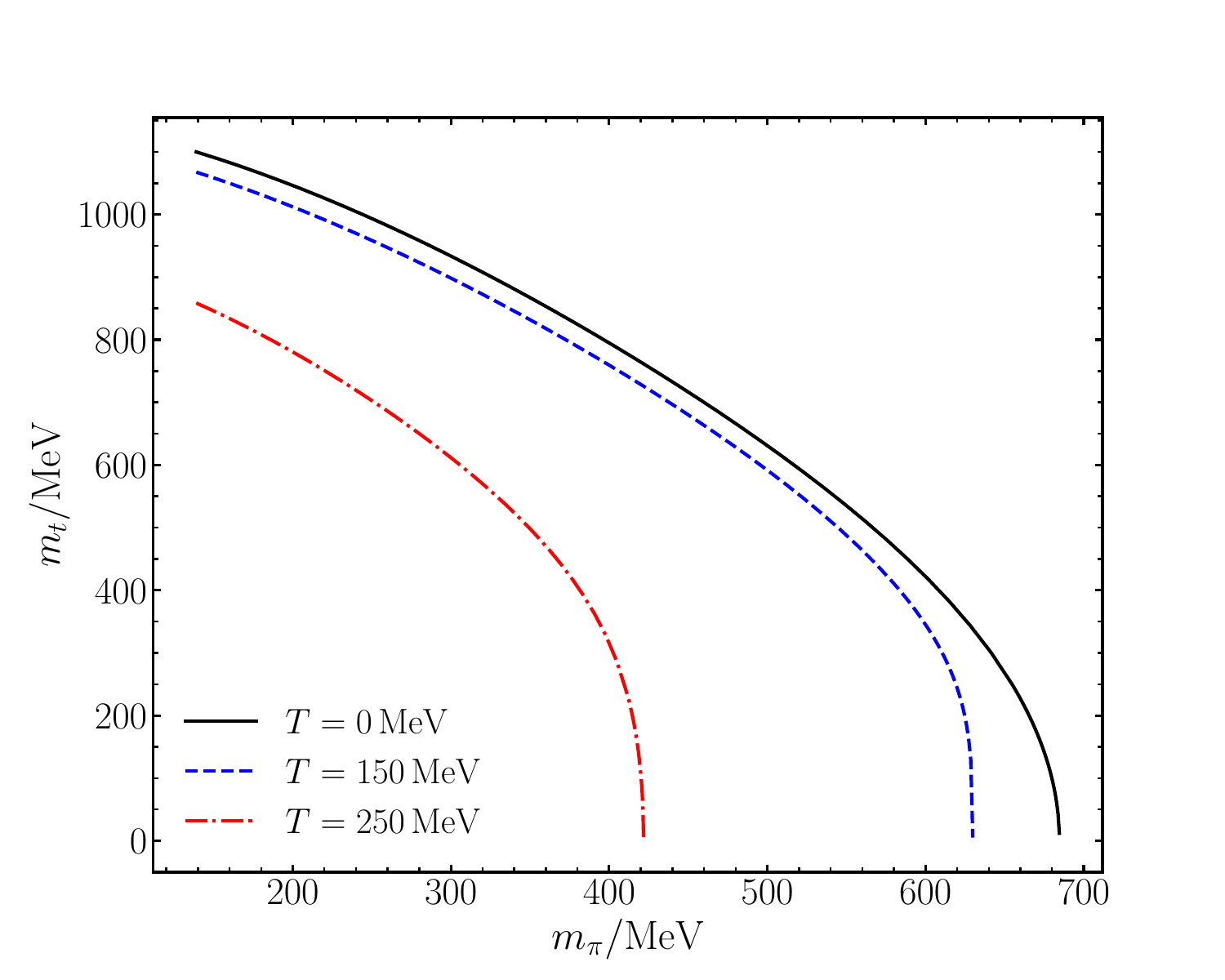}

    \caption{Left: tachyon pole positions with varying temperature for different zero-temperature pion masses ($m_\pi^I(0) = 200$, $139$ and $70$ MeV). Right: tachyon pole positions with varying zero-temperature pion masses for different temperatures ($T= 0$, $150$ and $250$ MeV). The varying $m_\pi$ is the zero-temperature value corresponding to the solution I of the gap equations~Eqs.~(\ref{eq:finiteT Gap eq 1}, \ref{eq:finiteT Gap eq 2}).  }   
    \label{fig:tachyon pos diff mpi and finiteT}   
\end{figure}
which is the finite temperature version of $i D^{-1}_{\tau\tau}$. The function $B^T(s,\chi,M)$ is the one-loop two-point integral encountered in the large $N$ calculation of the $O(N)$ model~\cite{Lyu:2024lzr} and can be decomposed into the zero-temperature part and the finite-temperature part, respectively~\cite{Bellac:2011kqa}: 
\begin{align} \label{eq:def of loop integral BT 1}
    B^{T}\left(s,\chi,M\right) &= B\left(s,\chi,M\right) + B^{T\neq 0}\left(s,\chi,T\right)\,, \\
    B^{T\neq 0}\left(s,\chi,T\right) &= \int^{\infty}_0 \frac{\mathrm d k \, k^2}{8\pi^2 \omega_k^2}
    n_B(\omega_k(\chi))\left(  \frac{1}{E+2\omega_k(\chi)} - \frac{1}{E-2\omega_k(\chi)} \right)\,,
    \label{eq:def of loop integral BT 2}
\end{align}
with $B\left(s,\chi,M\right)$ defined in Eq.~\eqref{eq: def of ren-Bfun} and $B^{T\neq 0}$ evaluated in the center-of-momentum (CM) frame and $s = E^2$. It is obvious that $B^{T\neq 0}$ is real and monotonically increasing in the $s<0$ region. Then with a similar argument as in the zero-temperature case, the function $iD^{-1}_{\tau\tau,T}(s)$ is monotonically increasing when $s<0$ and  $iD^{-1}_{\tau\tau,T}(s) \to -\infty$ as $s\to -\infty$. Thus the number of zero points of $iD^{-1}_{\tau\tau,T}(s)$ along $s<0$ axis, i.e. tachyons, cannot be more than one which is the same as the zero-temperature case. Similarly, the existence of a tachyon at finite temperature is determined by the sign of $iD^{-1}_{\tau\tau,T}(s)$ at $s=0$. The criterion for a tachyon to be present at finite temperature is given as
\begin{align}\label{eq:tachyon criterion}
   \Delta^T_t \equiv B^{T}(0,\chi,T)+\frac{v^2(T)/N}{\chi} >0\,.
\end{align}
The results for the existence of a tachyon  with varying temperature are illustrated in Fig.~\ref{fig:tachyon criterion finiteT} by showing the sign of the criterion function $\Delta^T_t$. Unlike the zero temperature case, the criterion for the
existence of a tachyon and the unstable vacuum condition no longer
agree with each other.\footnote{The origin of this difference is related to the subtlety at finite temperature that, in the CM frame, the inverse of the $\sigma$ propagator at zero-momentum, $[D^T_{\sigma\sigma}(\mathbf{p}=0,p^0\to 0)]^{-1}$, is not identical to $\partial^2 V^T/\partial \sigma^2|_{\sigma=0}$ for $T\neq 0$. }
Notably, for large unphysical pion masses and high temperatures the tachyon pole position $s_t = - m_t^2$ will move towards the positive real $s$-axis which is shown in Fig.~\ref{fig:tachyon pos diff mpi and finiteT}. The similar results for varying $M$ values are not plotted and we find the tachyon pole positions are pushed further away from the positive $s$-axis for larger $M$ values in both cases, i.e., with varying $m_\pi$ and temperature. Since the validity range of $O(N)$ model can be approximately taken to be $s\gg -m_t^2$ and $s-s_{th}(=4m_\pi^2(T)) \ll m_t^2$ for real $s$, the applicable energy range of $O(N)$ model is more and more limited as pion mass becomes larger and temperature goes higher.

\begin{figure}[!htbp]
    \centering
    \includegraphics[width=0.4\textwidth]{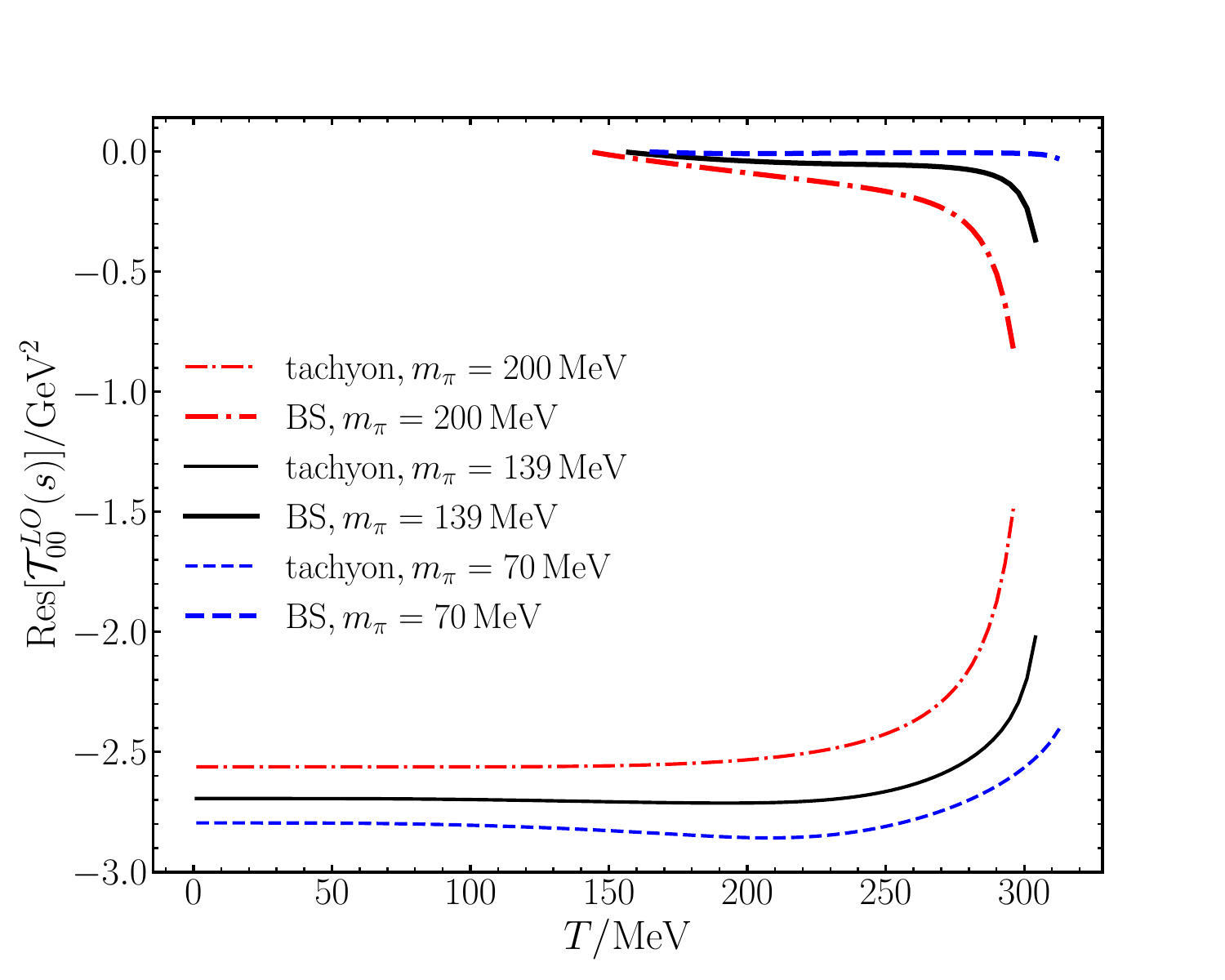}
    \includegraphics[width=0.4\textwidth]{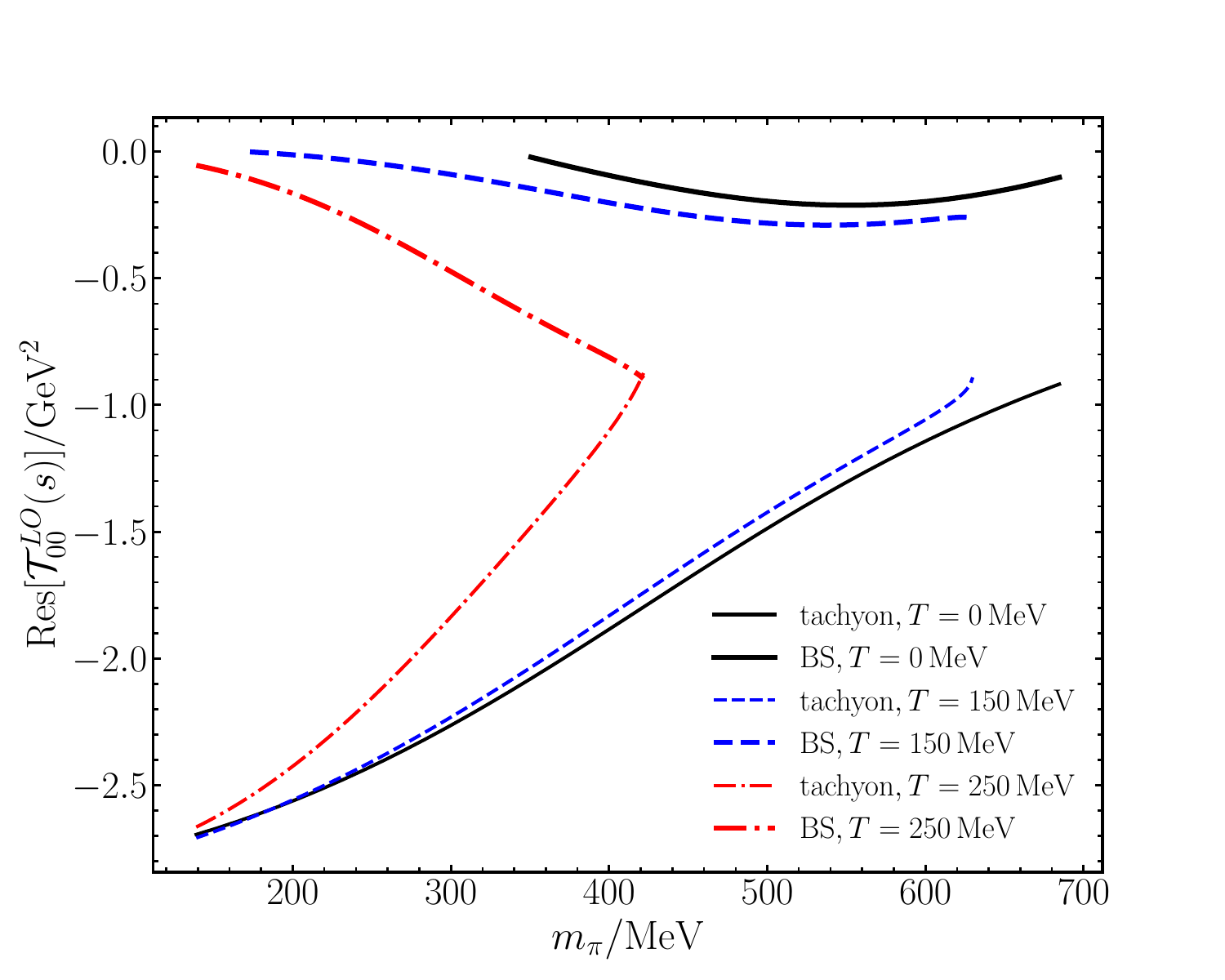}

    \caption{Residues of the leading $1/N$ order thermal amplitude $\mathcal T^{LO}_{00}$ at the tachyon pole and the $\sigma$ bound state (BS) pole. Left: residues with varying temperature for different pion masses ($m_\pi^I(0)=200$, $139$ and $70$ MeV). Right: residues with varying zero-temperature pion mass for different temperatures ($T=0$, $150$ and $250$ MeV).} 

    \label{fig:residues of the thermal amplitude}   
\end{figure}

It has been known since Heisenberg that the residue of $iS_{00}(k)$ for the
normal bound state is positive~\cite{Taylor:1972pty,
Newton:1982qc,Heisenberg:1946ytd,Hu:1948zz} in non-relativistic quantum
mechanics\footnote{We thank the referee for reminding us about this
interesting result.}. Though whether this result can be
generalized to relativistic quantum field theory is not quite clear,
to examine the residues of $S$ matrix at the tachyon pole and the bound
state is an interesting point. After investigating the residues for the tachyon pole and the bound
state pole numerically with varying temperature and pion mass, we find
that in the large $N$ limit of $O(N)$ model, the residues of
$\mathcal T_{00}(s)$ matrix with respect to $s$ for the tachyon pole
and the bound state pole are real and negative. See Fig.~\ref{fig:residues of the thermal amplitude} for an illustration.
However, since the relativistic $S_{00}$ matrix here is defined as
$S_{00}(s)=1+2i\rho(s) \mathcal T_{00}(s)$, due to the relativistic phase space
factor $\rho(s)=\sqrt{\frac {s-4m_\pi^2}s}$, the residue of $iS_{00}(k)$
 ( with respect to $k$, where $k=\sqrt{s-4m_\pi^2}/2$
is the momentum of pion in the CM frame)
for the tachyon pole will be purely imaginary because of an extra $i$
from $\sqrt s$ in $\rho(s)$ for $s<0$, whereas the residue for the
bound state is positive as usual since $s>0$. 
Since the $\mathcal T$ matrix does not have this $\sqrt s$ cut in the leading
$1/N$ approximation, the residues of $S(s)$ for the tachyon on the two
Riemann sheets on the $s$-plane defined by $\sqrt s$ cut would have
opposite signs (we suppose that the physical Riemann sheet of the $s$
plane includes the half real axis with angle $\pi$, while the
unphysical Riemann sheet includes the half real axis with angle
$-\pi$).  Thus, in this sense, the ``tachyon state" really can not be
regarded as a normal bound state.

\begin{figure}%[!htbp]
    \centering
    \includegraphics[width=0.4\textwidth]{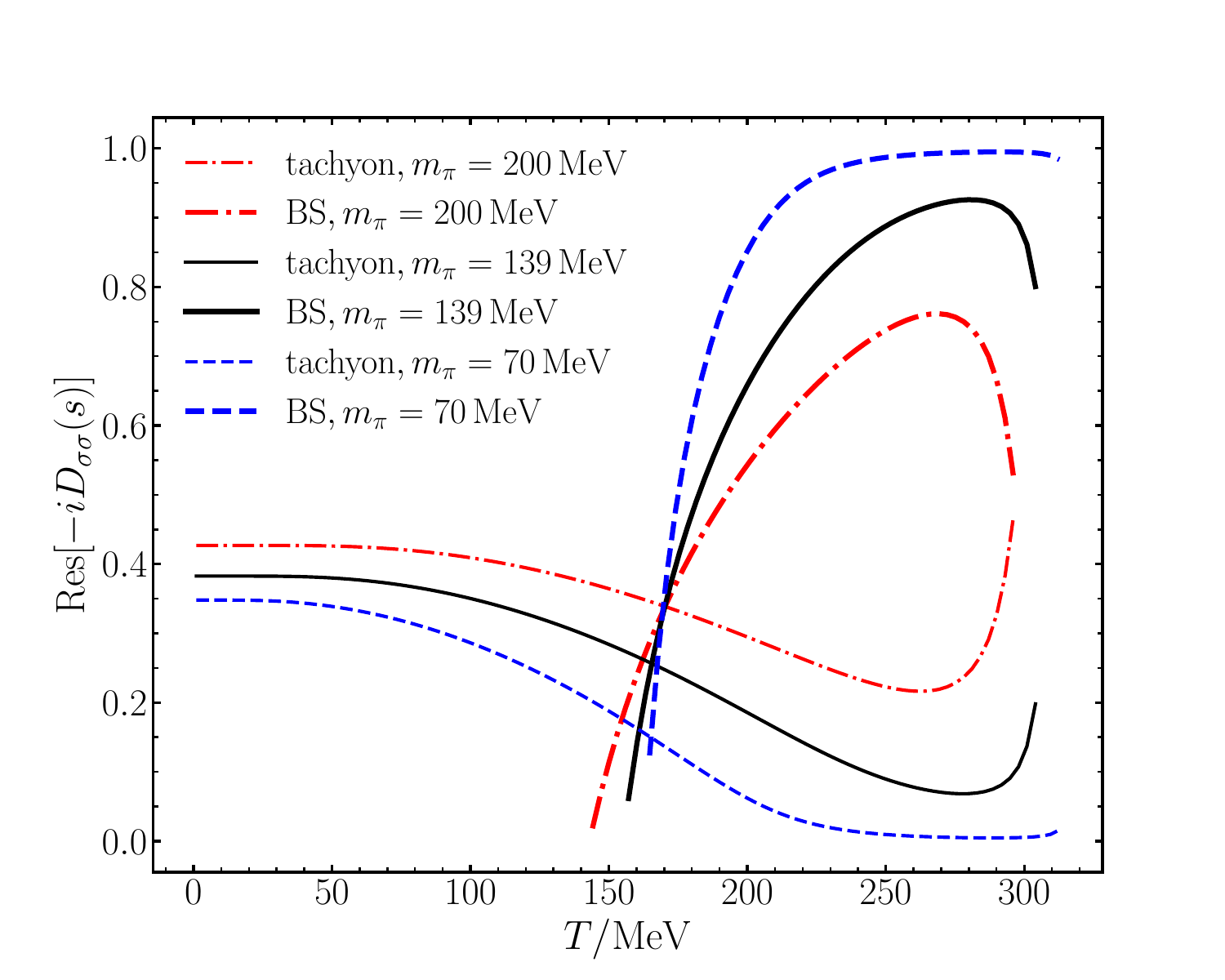}
    \includegraphics[width=0.4\textwidth]{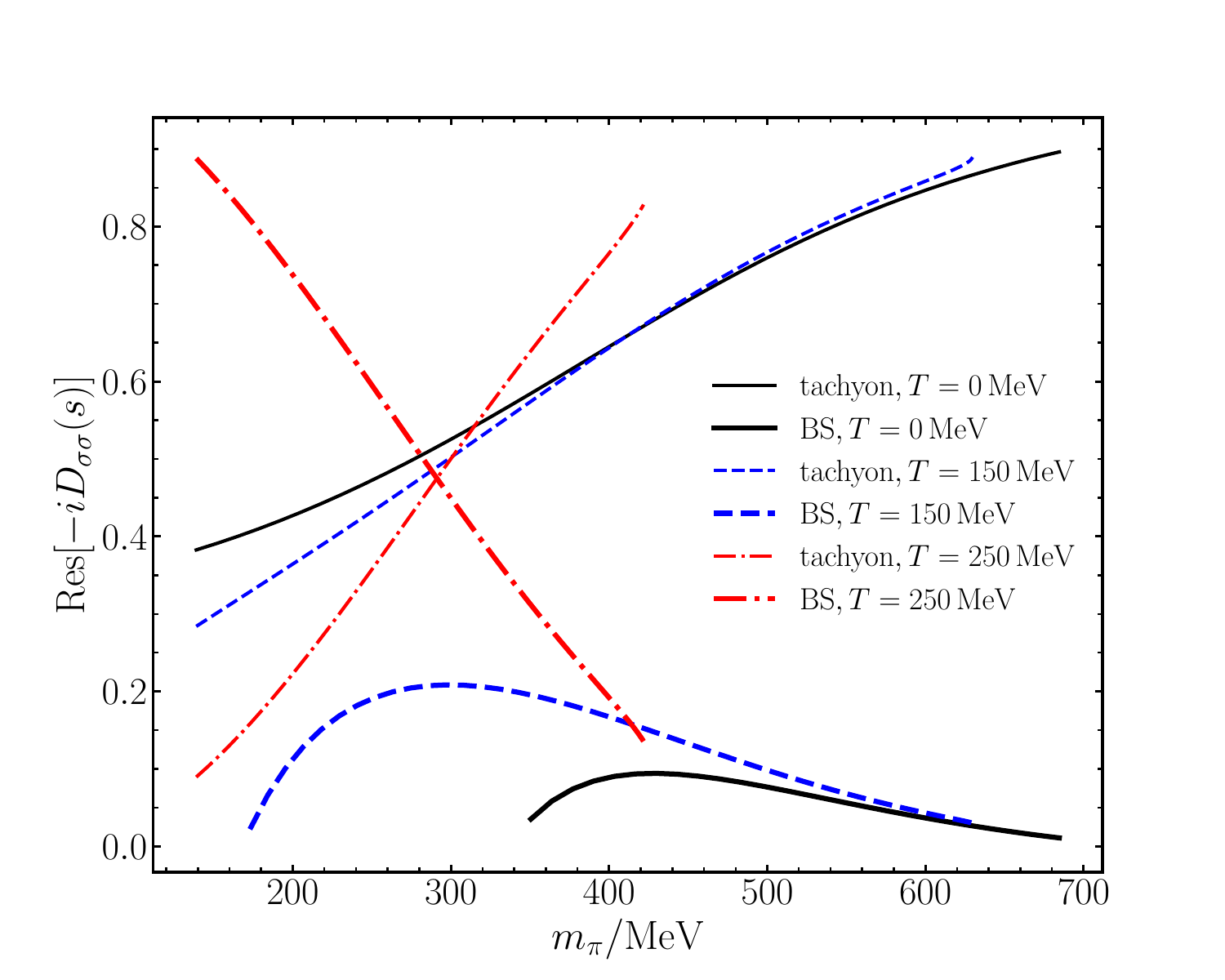}

    \caption{Residues of the finite-temperature $\sigma$ propagator in the leading $1/N$ order at the tachyon pole and the $\sigma$ bound state (BS) pole (for solution I of the gap equations~Eqs.~(\ref{eq:finiteT Gap eq 1}, \ref{eq:finiteT Gap eq 2})). Left: residues with varying temperature for different pion masses ($m_\pi^I(0)=200$, $139$ and $70$ MeV). Right: residues with varying zero-temperature pion mass for different temperatures ($T=0$, $150$ and $250$ MeV). The varying $m_\pi$ is the zero-temperature value corresponding to the solution I of the gap equations~Eqs.~(\ref{eq:finiteT Gap eq 1}, \ref{eq:finiteT Gap eq 2}).} 

    \label{fig:residues of the sigma propagator}   
\end{figure}

Additionally, it may also be instructive to examine if the residue at the tachyonic pole is positive for the finite-temperature $\sigma$ propagator in the leading order large $N$ expansion, which can be calculated in the CM frame as~\cite{Coleman:1974jh,Lyu:2024lzr},
\begin{align}
     D_{\sigma\sigma}(s)=\frac{i B^{T}(s,\chi,M)}{(s-\chi) B^{T}(s,\chi,M) - v^2(T)/N}.
\end{align} 
By comparing $i D^{-1}_{\tau\tau,T}$ and the denominator of
$D_{\sigma\sigma}$, it is obvious that  the $\sigma$ propagator and
the $\pi\pi$ thermal amplitude $\mathcal T^T_{00}$ have the  same pole
positions but with different residues. The residues of $-i
D_{\sigma\sigma}$ at both the bound state pole and the tachyon pole
are shown in Fig.~\ref{fig:residues of the sigma propagator} with
varying temperature and pion mass. It is found that the tachyonic pole
in the $\sigma$ propagator has a positive definite residue as the
$\sigma$ bound state pole. %, which means that in the $O(N)$ model, the tachyon does not lead to negative norm state in the Hilbert space of the system. 
However, the tachyon still should not be put on an equal footing with
a usual bound state, because the existance of a tachyon may actually
demonstrate that the present chosen vacuum is not the absolute
minimum of the effective
potential~\cite{Coleman:1974jh,Abbott:1975bn,Linde:1976qh}.

\section{The double-branch effective potential and the two phases of $O(N)$ model\label{sect:eff_coupling}}

The double-branch feature of the effective potential of the $O(N)$ model at large $N$ limit has been known for a long time~\cite{Kobayashi:1975ev,Abbott:1975bn,Linde:1976qh,Bardeen:1983st,Bardeen:1986td,Nunes:1993bk}. However, besides the mathematical results, not so much physics has been understood about the nature of the two branches of $V(\phi)$. For zero temperature and without explicit symmetry breaking, an interesting discovery in Ref.~\cite{Linde:1976qh} states that the effective coupling constant obtained from the effective potential has opposite signs on the two branches, i.e., the effective coupling constant is positive on the first branch and negative on the second.  In the following, 
we reinvestigate this correspondence for nonzero explicit symmetry breaking and finite temperature.

\begin{figure}
    \centering
    \includegraphics[width=0.4\textwidth]{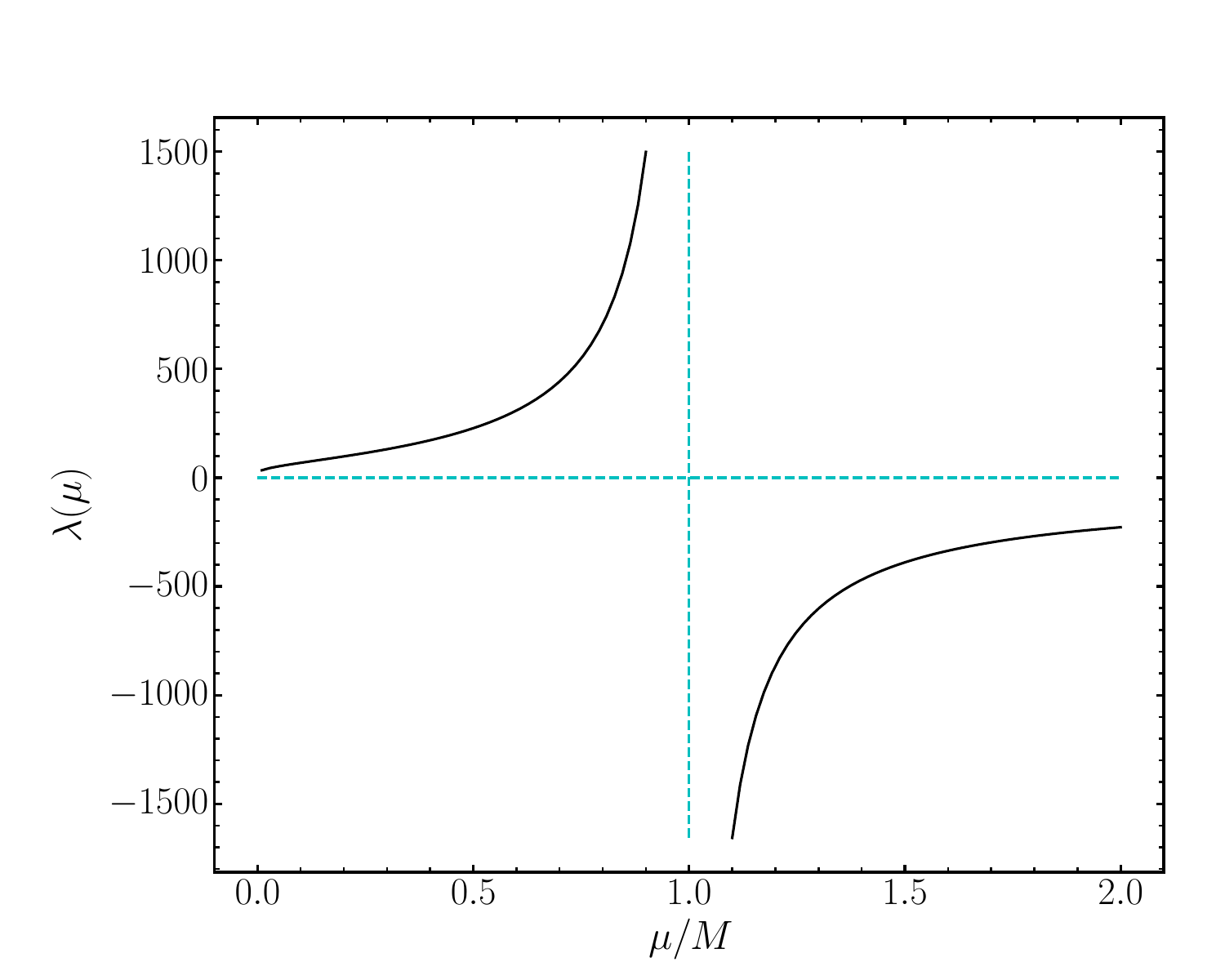}

    \caption{The renormalized coupling constant $\lambda(\mu)$. The result is obtained within leading order large $N$ expansion as in Eq.~\eqref{eq:renormalized lambda of mu}.}
    \label{fig:renormalized coupling}    
\end{figure}

\begin{figure}%[!htbp]       
    \centering
    \includegraphics[width=0.4\textwidth]{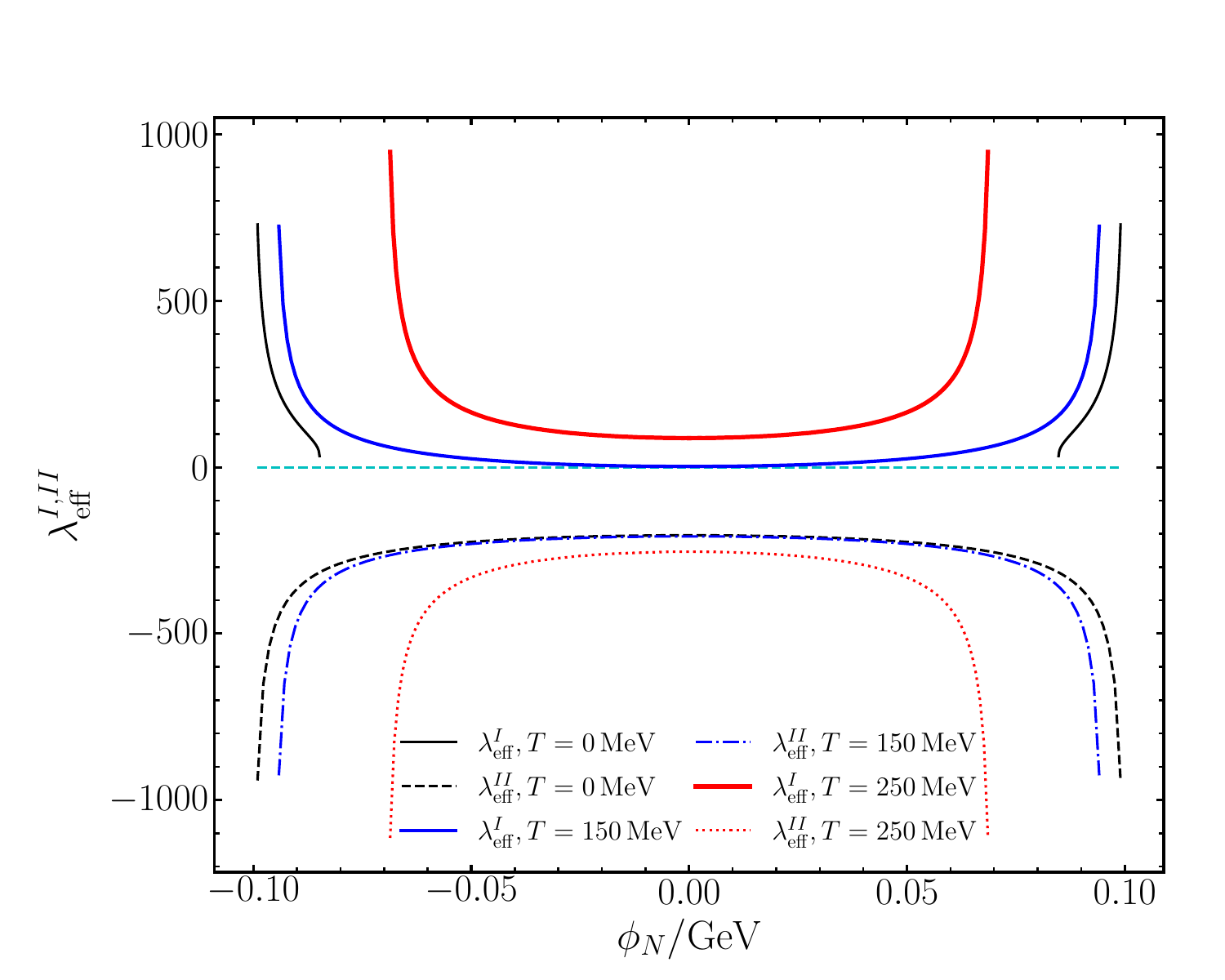}

    \caption{Effective coupling constant for the two branches of the effective potential. The illustrated cases are $T=0$, $150$ and $250$ MeV respectively.
    For the first branch, $\lambda^I_\text{eff}>0$ and for the second branch, $\lambda^{II}_\text{eff}<0$, with the boundary between the two branches, i.e., $\phi^2=\phi_{b,T}^2$, corresponding exactly to the Landau pole. 
    The correspondence between the two branches of the effective potential and the two phases (positive and negative coupling) of $O(N)$ model is verified with nonzero explicit symmetry breaking and finite temperature. }
    \label{fig:eff coupling}    
\end{figure}

As usual, the effective coupling  constant $\lambda_\text{eff}(T)$ can be defined as
\begin{align}
    \left.\frac{\partial^4 V^T}{\partial \pi_a^4}\right|_{\pi_a=0} \equiv \frac{3}{N} \lambda_\text{eff}(T)\,.
\end{align} 
At zero temperature, the effective coupling can be exactly calculated,
\begin{align}\label{eq:lambda eff zeroT}
    \lambda_\text{eff}(T=0) = \frac{32\pi^2}{\log(M^2/\chi(v^2))-1} = \lambda\left(\mu =\sqrt{e\chi(v^2)} \right)\,,
\end{align}
where $\chi(v^2)$ is solved from Eq.~\eqref{eq:zeroT Gap eq 1} which has two solutions of $\chi$ corresponding to the two branches respectively and $\lambda(\mu =\sqrt{e\chi(v^2)})$ is the renormalized coupling constant $\lambda(\mu)$ evaluated at the scale $\mu = \sqrt{e\chi(v^2)}$. $\lambda(\mu)$ can be obtained from the renormalization condition Eq.~\eqref{eq:ren-cond2}, and the result is simply expressed as
\begin{align}\label{eq:renormalized lambda of mu}
    \lambda(\mu) = \frac{32\pi^2}{\log (M^2/\mu^2)}\,,
\end{align}
with $M$ the intrinsic scale of $O(N)$ model where the coupling constant diverges, i.e. the Landau pole. The sketch of the renormalized coupling constant is shown in Fig.~\ref{fig:renormalized coupling}. It is obvious that for $\mu<M$, $\lambda(\mu)>0$ and for $\mu>M$, $\lambda(\mu)<0$. 
In fact, it can also be seen from Eq.~\eqref{eq:lambda eff zeroT} that $\lambda_\text{eff} >0$ for the first branch and $\lambda_\text{eff} <0$ for the second branch, whereas the boundary between the two branches, i.e., $\phi^2=\phi_b^2 = \phi^2(\chi=M^2/e)$, corresponds exactly to the Landau pole at $\mu=M$. In other words, the two branches of the effective potential still correspond to the two branches of the renormalized coupling constant, with $V^{I}(\phi)$ corresponding to the positive coupling branch whereas $V^{II}(\phi)$ corresponding to the negative coupling branch.
At finite temperature, the effective coupling constant\footnote{In fact, the effective coupling constant can also be obtained considering that the effective potential is the generating function for one-particle-irreducible (1PI) Green's functions with all external momenta ($p^\mu_i$) set to zero. At finite temperature, the only subtlety is that we have to first set $p^0_i=0$ and then take the limit $|\mathbf{p}_i| \to 0$, since the 1PI finite-temperature Green's functions may not be analytic at this point due to the Landau cut~\cite{Weldon:1983jn,das1997finite} and the effective potential is a quantity related to the static properties of the system, which is in the thermal equilibrium.} can be calculated as,
\begin{align}
    \lambda_\text{eff}(T) = \frac{32\pi^2}{\log(M^2/\chi)-1 + 4\int^\infty_0 \mathrm d k \, n_B(\omega_k(\chi))/\omega_k(\chi) } = 2N\left( \frac{\mathrm d}{\mathrm d \chi} \phi^2(\chi)\right)^{-1}\,,
\end{align}
with $\chi = \chi(v^2)$ as mentioned above and $\phi^2(\chi)$ defined in Eq.~\eqref{eq:finiteT Gap eq 1}. Noticing that at the boundary between the two branches, $\phi^2=\phi^2_{b,T}=\phi^2(\chi = \chi_{b,T})$, which is the maximum of $\phi^2(\chi)$ with  $\frac{\mathrm d}{\mathrm d \chi} \phi^2(\chi)=0$, we have $\lambda_\text{eff}(T)=\infty$, corresponding to the Landau pole. Also, it can be seen that $\frac{\mathrm d^2}{\mathrm d \chi^2} \phi^2(\chi)<0$, then for the first branch ($\chi<\chi_{b,T}$), $\frac{\mathrm d}{\mathrm d \chi} \phi^2(\chi)>0$ such that $\lambda_\text{eff}>0$; for the second branch ($\chi>\chi_{b,T}$), $\frac{\mathrm d}{\mathrm d \chi} \phi^2(\chi)<0$ then $\lambda_\text{eff}<0$. 
Therefore, at finite temperature, the correspondence between the two branches of the effective potential and the two phases of $O(N)$ model still holds, which is also illustrated in Fig.~\ref{fig:eff coupling}. 

The above results are actually consistent with a recent argument in Ref.~\cite{Romatschke:2023sce} that $O(N)$ model is not trivial even with infinite UV cutoff, if the negative coupling phase is acceptable. Remarkably, the negative coupling does not make the vacuum unstable, which has been clearly seen in Refs.~\cite{Kobayashi:1975ev,Abbott:1975bn,Linde:1976qh,Nunes:1993bk} when there is no explicit breaking. Instead, the vacuum is stabilized by the non-perturbative effects captured  in the large $N$ expansion. It has been shown in Sec.~\ref{sect:Veff_zeroT} and \ref{sect:Veff_finiteT} that with massive pions for the first branch and at finite temperature, the second branch of effective potential always has a local minimum with lower energy. Notably, in the light of the aforementioned correspondence, we may take a more physical view of the origin of the two branches: there are two branches of the effective potential generated by non-perturbative effects, because in the non-perturbative regime there are two phases of the $O(N)$ model, with positive or negative coupling. 

\section{Thermal $\sigma$ pole  trajectory in $N/D$ modified $O(N)$ model \label{sect:sigma_traj}}

With the $N/D$ method, it has been shown in Ref.~\cite{Lyu:2024lzr} that at zero temperature, the crossing symmetry of $O(N)$ model in the large $N$ limit could be partially recovered while preserving unitarity. In fact, a similar procedure can also be applied to $O(N)$ model with finite temperature and can offer a more complete knowledge of the thermal properties of the $\sigma$ particle, after taking account of the thermal contribution from the crossed channels. 

Similar to the case of $O(N)$ model at zero temperature~\cite{Lyu:2024lzr}, we can add the $t$- and $u$-channel contributions to partially recover crossing symmetry.  After isospin decomposition and partial wave projection, the $IJ=00$ channel thermal amplitude is expressed as
\begin{align}\label{eq:ONamp cross channel finiteT}
    \mathcal T_{00}^T(s) = \frac{N-1}{32\pi} \mathcal A_{LO}^T(s) + I_{tu}^T(s) \,,
\end{align}
where
\begin{align}
    \mathcal A_{LO}^T(s) &=  \frac{m_\pi^2(T) - s}{ (s - m_\pi^2(T)) N B^T(s,m^2_\pi(T),M)
- v^2(T)}\,, \\
    I_{tu}^T(s) &= \frac{1}{16\pi (s - 4m_\pi^2(T))} \int_{4m_\pi^2(T) -
s}^0 \mathrm d t' \mathcal A_{LO,tu}^T(t')\,,
    \label{eq:ONamp Itu finiteT} \\
    \mathcal A_{LO,tu}^T(t) &=  \frac{m_\pi^2(T) - t}{ (t - m_\pi^2(T)) N B^T_{tu}(t,m^2_\pi(T),M)- v^2(T)}\,, 
\end{align}
with   $m_\pi(T)$ and $v(T)$ solved from the leading $1/N$ order gap equations Eqs.~(\ref{eq:finiteT Gap eq 1}, \ref{eq:finiteT Gap eq 2}) on the first branch of the effective potential, and $B^T(s,m^2_\pi(T),M)$ defined as Eqs.~(\ref{eq:def of loop integral BT 1}, \ref{eq:def of loop integral BT 2}). $I_{tu}^T(s)$ is the $IJ=00$ channel partial wave projection integral of the cross-channel thermal amplitudes. The definition of $B^T_{tu}(t,m^2_\pi(T),M)$ appearing in the cross-channel thermal amplitude $\mathcal A_{LO,tu}^T(t)$ will be demonstrated as follows. Since at finite temperature, Lorentz invariance is not preserved anymore, things become more complicated  when dealing with the cross-channel thermal loop integral $B^{T}_{tu}$.  In fact, the thermal loop integrals depend differently on the temporal and spatial components of the external momentum. As a result, crossing symmetry is also broken and for $\pi\pi$ scattering the contributions from $t$- and $u$-channel are the same~\cite{Dobado:2002xf,GomezNicola:2002tn}:
\begin{align}
    B^{T}_{tu}\left(t,m_\pi(T),M\right) &= B\left(t,m_\pi(T),M\right) + B^{T\neq 0}_{tu}\left(t,m_\pi(T)\right)\,, \\
    B^{T\neq 0}_{tu}\left(t,m_\pi(T)\right) &= \left\{
    \begin{aligned}
    &\frac{1}{8\pi^2\sqrt{-t}}\int^\infty_0 \frac{\mathrm d k \, k}{\omega_k} n_B(\omega_k) \log \left|  \frac{\sqrt{-t} + 2k}{\sqrt{-t}-2k} \right|\,, 
    & &(t<0) \,,\\
    &\frac{1}{8\pi^2\sqrt{-t}}\left[
    \int^\infty_0 \frac{\mathrm d k \, k}{\omega_k} n_B(\omega_k) \log   \frac{\sqrt{-t} + 2k}{\sqrt{-t}-2k}\right. \\
    &\left.\phantom{\frac{1}{8\pi^2\sqrt{-t}}[} + i \pi T \mathrm{sign}(\operatorname{Im} t) 
    \log\left(1- e^{-\beta\sqrt{m^2_\pi(T)-t/4}}  \right) \right]
    \,, 
    & &(\operatorname{Im} t \neq 0) \,.
    \end{aligned}
    \right.
\end{align}
It is worth mentioning that the calculations of the thermal amplitudes are all conducted in the CM frame. 

In general, considering only the two-particle intermediate
states, the finite temperature unitarity for the  $IJ=00$ amplitude can
be deduced~\cite{Dobado:2002xf,GomezNicola:2002tn,GomezNicola:2023rqi},
\begin{align}\label{eq:thermal unitarity}
    \operatorname{Im} \mathcal T^{T}(s) = \rho^T(s)\left|\mathcal T^{T}\right|^2 \,, 
\end{align}
with $\rho^T(s) \equiv \rho(s)\left(1+2n_B(\sqrt{s}/2)\right)$ where
$\rho(s) = \sqrt{1-4m_\pi^2(T)/s}$. It can be verified that the leading order amplitude $N/(32\pi)\mathcal
A^T_{LO}(s)$  alone satisfies this relation.
But Eq.~\eqref{eq:ONamp cross channel finiteT} incorporating the $t$- and $u$-channel contributions which are at $\mathcal O(1/N)$, breaks the above relation. In pursuit of recovering the thermal unitarity and including the crossed channel contributions, the $N/D$ method can be adapted accordingly to studying thermal properties of the $\sigma$ particle within $N/D$ modified $O(N)$ model. The thermal amplitude can be expressed as
\begin{align}
    \mathcal T^T(s) = \frac{N^T(s)}{D^T(s)}\,,
\end{align}
with $N^T(s)$ containing only the left-hand cut ($L$) and $D^T(s)$ containing only the right-hand cut ($R$), i.e. the thermal unitary cut. In the light of the thermal unitarity relation Eq.~\eqref{eq:thermal unitarity}, we can obtain the relations between $N^T(s)$ and $D^T(s)$:
\begin{align}
    \operatorname{Im}_R D^T(s) &= -\rho^T(s)N^T(s)\,, \\
    \operatorname{Im}_L N^T(s) &= D^T(s)\operatorname{Im}_L \mathcal T^T(s)\,.
\end{align}
It is straightforward to write down dispersion relations for $N^T(s)$ and $D^T(s)$ by using the Cauchy integral. For the same reason~\cite{Lyu:2024lzr}, we still use \emph{twice subtracted} dispersion relations as in the zero-temperature case for consistency. The dispersion relations for $N^T(s)$ and $D^T(s)$ can then be written as
\begin{align}\label{eq: N/D twice sub for D finiteT}
        D^T(s) &= \frac{s-s_A}{s_0-s_A}  + g_D^T\frac{s-s_0}{s_A-s_0} - \frac{(s-s_0)(s-s_A)}{\pi} \int_R \frac{\rho^T(s')N^T(s')}{(s'-s)(s'-s_0)(s'-s_A)} 
        \mathrm d s' \,,\\
        N^T(s) &=b_0^T \frac{s-s_A}{s_0-s_A} + g_N^T\frac{s-s_0}{s_A-s_0} + \frac{(s-s_0)(s-s_A)}{\pi} \int_L \frac{D^T(s')\operatorname{Im}_L \mathcal T^T(s')}{(s'-s)(s'-s_0)
        (s'-s_A)} \mathrm d s' 
        \label{eq: N/D twice sub for N finiteT}\,,
\end{align}
where the subtraction points $s_A = m_\pi^2(T)$, $s_0 = s_{th}=4m_\pi^2(T)$, and $D^T(s_0)=1$ are chosen for convenience. With these choices, the subtraction constants are $N^T(s_0) = b_0^T$, $N^T(s_A) = g_N^T$, $D^T(s_A) = g_D^T$. The $\operatorname{Im}_L \mathcal T^T$ is extracted from Eq.~\eqref{eq:ONamp cross channel finiteT} as part of the $O(N)$ model input. 
Then the functions $N^T(s)$ and $D^T(s)$ can be solved numerically after setting the subtraction constants. 
\begin{figure}%[!htbp]
    \centering
    \includegraphics[width=0.4\textwidth]{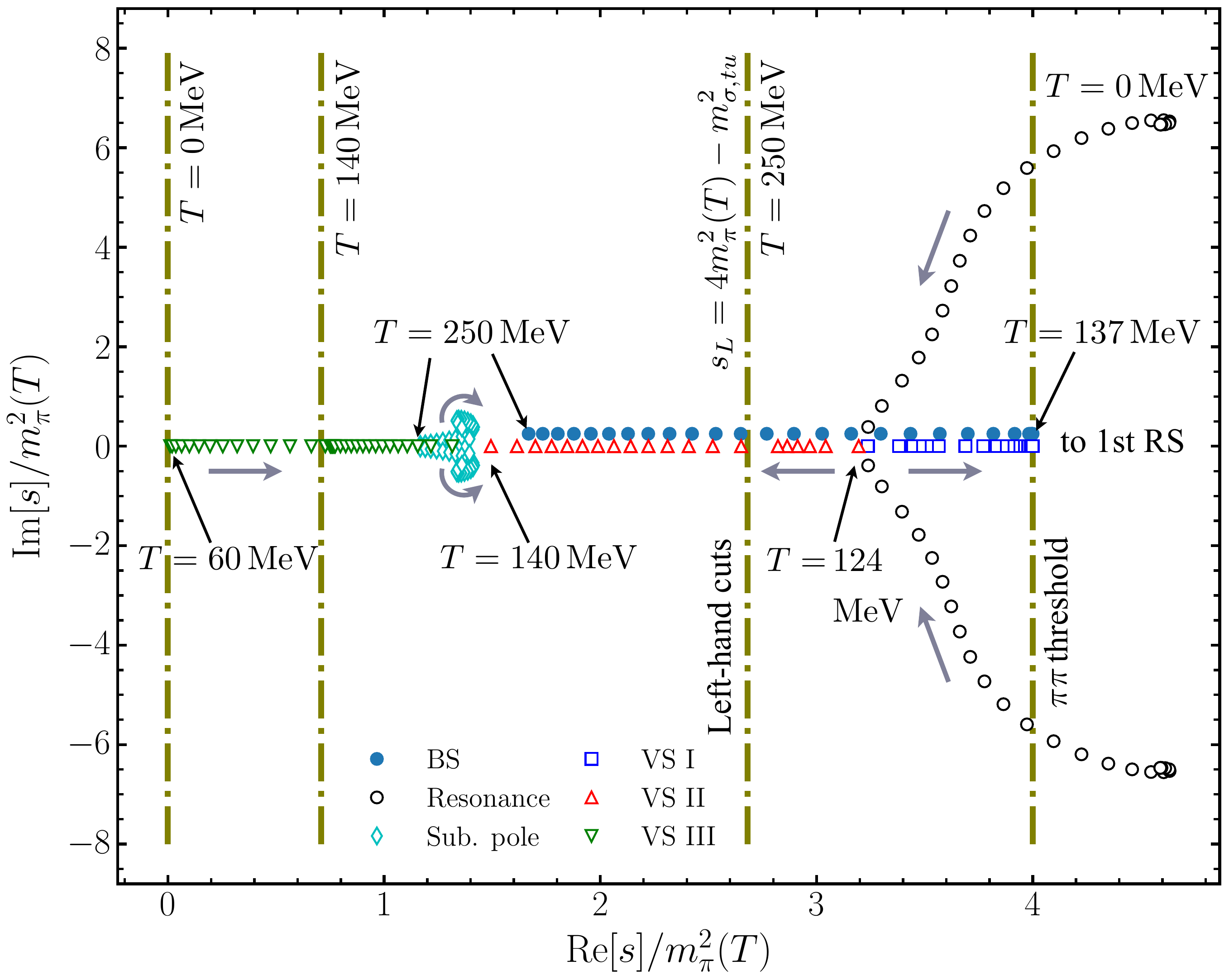}

    \caption{The thermal $\sigma$ pole  trajectory obtained in $N/D$ modified $O(N)$ model. The zero-temperature pion mass is set as $m_\pi(0)=139\,\mathrm{MeV}$. When the temperature increases, similar to the leading order result~\cite{Lyu:2024lzr}, $\sigma$ turns into two virtual states (VS I\&II) and then becomes a bound state (BS) after VS I moves towards and across the threshold to the first Riemann sheet (RS). The left-hand cut branch point extends to $s_L = 4m_\pi^2(T) - m^2_{\sigma,tu}$ due to the $\sigma$ exchange in crossed channels. Additionally, the third virtual state pole (VS III) generated close to $s_L$ will meet VS II on the real axis, then becoming a pair of subthreshold (Sub.) poles and going into the complex plane. Finally, the pair of subthreshold poles tends to $s_A = m_\pi^2(T)$ as $T \gg T_c$.  } 
    \label{fig:finiteT_ND_traj}   
\end{figure} 
According to the prescription chosen for the $N/D$ modified $O(N)$ model at zero temperature~\cite{Lyu:2024lzr}, with finite temperature the subtraction constants are determined consistently to be
\begin{align}
    b_0^T &= \mathcal T_{00}^T(s_0) = \frac{N-1}{32\pi} \mathcal A^T_{LO}(s_0) + I^T_{tu}(s_0)\,, \label{eq:Constraint1}\\
    g_D^T &=\frac{32\pi v^2(T) b_0^T}{N(s_0-s_A)}\,, \label{eq:Constraint2}\\
    \frac{g_N^T}{g_D^T} &= \operatorname{Re} I^T_{tu}(s_A)\,. \label{eq:Constraint3}
\end{align}
When ignoring the  $\mathcal O(1/N)$ contributions, the above choice of subtraction constants can exactly recover the leading $1/N$ order thermal amplitude of the $O(N)$ model. The main difference between the finite temperature and the zero-temperature prescriptions is that for high temperatures such that $\sigma$ becomes a bound state, there is no need to restrict the bound state pole positions in $s$-channel and crossed channels to be the same, due to the breaking of crossing symmetry caused by the thermal loop integrals. In addition, the intrinsic scale is now set with a different value $M = 1.5$ GeV (in consistency with Ref.~\cite{Lyu:2024lzr}) due to the higher order contributions incorporated through the $N/D$ unitarization of the $O(N)$ model scattering amplitude.

The numerical results of the thermal sigma pole trajectory is depicted in Fig.~\ref{fig:finiteT_ND_traj}. When temperature increases,  the broad $\sigma$ resonance turns into two virtual states (VS I and II) first and then the upper virtual state, VS I, moves towards and through the threshold to the first Riemann sheet and becomes a bound state. For $T\gg T_c (= \sqrt{12/N} f_\pi \simeq 160 \,\text{MeV})$, the $\sigma$ bound state pole $s_\sigma$ moves close to $m^2_\pi(T)$, i.e., $\sigma$ tends to become asymptotically degenerate with pions, as expected by the chiral symmetry restoration. This behaviour is similar to the leading $1/N$ order calculations in 
Refs.~\cite{Lyu:2024lzr,Patkos:2002xb,Patkos:2002vr}~\footnote{After the present work was completed, Z. Sz\'ep notified us of the related earlier works, which we did not notice before. We are grateful to Z. Sz\'ep for pointing out these works to us. 
In Ref.~\cite{Patkos:2002vr}, the thermal $\sigma$ pole trajectory within the leading $1/N$ order calculation of the $O(N)$ linear $\sigma$ model was also obtained, where a version of cutoff regularization is adopted and the $\sigma$ pole position is determined from the $\sigma$ propagator. This $\sigma$ pole trajectory with varying temperature is qualitatively consistent with the results in our previous paper Ref.~\cite{Lyu:2024lzr}, whereas the large $N$ results are very different from those calculated in Ref.~\cite{Hidaka:2002xv} within the so-called ``optimized perturbation theory" resummation scheme.}, however the main differences are  as follows. With cross-channel thermal contributions, the left-hand cut extends to $(-\infty, s_L)$ with $s_L = 4m_\pi^2(T) - m^2_{\sigma,tu}$ where $m_{\sigma,tu}$ is determined by the $\sigma$ bound state pole positions in the $t$- and $u$-channel thermal amplitudes, which are identical for $\pi\pi$ scattering. Additionally, there is a third virtual state (VS III) generated close to $s_L$. For higher temperatures, VS II and III move towards and then hit each other, becoming a pair of subthreshold  poles, which are similar to those also found in Refs.~\cite{Cao:2023ntr,Lyu:2024lzr} for large unphysical pion mass at zero temperature. 
Finally, the subthreshold poles move close to $s_A = m_\pi^2(T)$ as $T \gg T_c$. This results directly from the fact that, on the second Riemann sheet $D^{T,II}(s) = D^T(s) + 2i\rho^T(s) N^T(s)$, thus $D^{T,II}(s_A) \to 0$ as $v(T) \to 0$ for high temperatures with the subtraction constants set as Eqs.~(\ref{eq:Constraint2}, \ref{eq:Constraint3}).  

To one's surprise, the thermal $\sigma$ pole trajectory demonstrates again analogous structure and behavior with the pole trajectory  with varying $m_\pi$ at zero temperature. Actually, this phenomenon is understandable since the third virtual state, VS III, and the subthreshold poles are generated mainly by the interplay of (thermal) unitarity, the behavior of the (thermal) amplitude zero point (i.e. Adler zero for zero temperature), and the cross-channel effects~\cite{Cao:2023ntr,Lyu:2024lzr}.

\section{Conclusions and discussions\label{sect:conclusion}}

In this work, we revisit the leading $1/N$ order effective potential of the $O(N)$ linear $\sigma$ model with explicit symmetry breaking. After a careful investigation for the cases with different pion masses and temperatures, a new phenomenon is found for large $m_\pi$ values and high temperatures: as either $m_\pi$ or temperature increases, the local minimum on the first branch of the effective potential,  which is the phenomenologically preferred vacuum, will move to the second branch and turn into a saddle point. The stable vacuum  always lies on the second branch. However,  it is not preferred phenomenologically,  since it does not have spontaneous chiral symmetry breaking in the chiral limit.  

As for the tachyon pole, at larger $m_\pi$ and higher temperature, its position moves towards the positive real axis, causing the validity domain of $O(N)$ model to shrink (when the vacuum is set as the first-branch solution of the gap equations Eqs.~(\ref{eq:finiteT Gap eq 1}, \ref{eq:finiteT Gap eq 2})). 
However in the $O(N)$ model, it seems that even for high temperatures $T_c(\sim 160\,\text{MeV}) < T \lesssim 300$ MeV \footnote{For the leading $1/N$ order calculation and with the intrinsic scale set as $M=550$ MeV.} and before the whole effective theory breaks down, there is still a ``window'' for the $\sigma$  particle to survive and become asymptotically degenerate with pions.
Additionally, even with nonzero explicit symmetry breaking and at finite temperature, we find the possible correspondence~\cite{Linde:1976qh} between the two branches of the  effective potential and the two phases (positive and negative coupling) of the $O(N)$ model still holds, with the help of the effective coupling constant.

As the completion to our previous work in Ref.~\cite{Lyu:2024lzr}, the $N/D$ method is generalized to unitarize the amplitude at finite temperature while partially including the cross-channel contribution. With this thermal amplitude of the $N/D$ modified $O(N)$ model,  the $\sigma$ pole trajectory at finite temperature is also studied. Similar to the case with varying pion mass at zero temperature, at high temperatures, the thermal $\sigma$ pole becomes a bound state pole on the first sheet, and a pair of resonance poles far below the threshold is also found, which is related to the cross-channel contribution.

One of the possible directions for future study could be the effects on the vacuum structure of $O(N)$ model of nonzero $m_\pi$ and finite temperatures when high dimensional terms and higher $1/N$ order contributions are included. Additionally, it is possible to explore the vacuum structure of $O(N)$ model even in curved spacetime such as (anti-)de Sitter space or Rindler space~\cite{Serreau:2011fu,Gautier:2015pca,LopezNacir:2020dik,Basu:2023bcu}.\footnote{The authors thank S. R. Haridev for pointing us to recent interesting works on the $O(N)$ model in curved spacetime.} Moreover, the lessons learned in $O(N)$ model may arouse the interest in the structure of the QCD vacuum at large pion masses and high temperatures.

\begin{acknowledgments}
This work is supported by China National Natural Science Foundation
under Contract No. 12335002, 12375078, 11975028.
This work is also supported by “the Fundamental Research Funds for the Central Universities”.
\end{acknowledgments}

\bibliographystyle{apsrev4-2}
\bibliography{ONVeff}

\end{document}